\documentclass[12pt,a4paper,english]{article}
\usepackage{titling}
\usepackage[affil-it]{authblk}
\usepackage[dvipsnames]{xcolor}
\usepackage{longtable}
\usepackage{lscape}
\usepackage{graphicx}
\usepackage{epsfig}
\usepackage{dsfont}
\usepackage{amsfonts} 
\usepackage{amsmath}
\usepackage{amsthm}
\usepackage{amssymb}
\usepackage{bbold}
\usepackage[english]{babel}
\usepackage[utf8]{inputenc}
\usepackage{calc}
\usepackage{cancel}
\usepackage{float}
\usepackage{slashed}
\usepackage{mathrsfs}
\usepackage{pdfpages}
\usepackage{tabu}
\usepackage{simplewick}
\usepackage[hidelinks]{hyperref}
\usepackage{multirow}
\usepackage{hhline}
\usepackage{float}
\usepackage{bm}
\usepackage{braket}
\usepackage{comment}
\usepackage{cite}
\usepackage[left=0.8in,right=0.8in,top=1.0in,bottom=1.2in]{geometry}

\usepackage{datetime}
\usepackage{booktabs}
\date{}

\numberwithin{equation}{section}
\numberwithin{figure}{section}
\numberwithin{table}{section}

\makeatletter
\g@addto@macro\bfseries{\boldmath}
\makeatother

\allowdisplaybreaks

\newcommand{\published}[1]{%
\gdef\puB{#1}}
\newcommand{\puB}{}

\newcommand{\be} {\begin{equation}}
\newcommand{\ee} {\end{equation}}
\newcommand{\bea} {\begin{eqnarray}}
\newcommand{\eea} {\end{eqnarray}}
\newcommand{\no} {\nonumber}

\newcommand{\cR}{{\mathcal R}}

\newcommand{\cL}{{\mathcal L}}
\newcommand{\cH}{{\mathcal H}}
\newcommand{\tcH}{\widetilde{\mathcal H}}

\newcommand{\cA}{{\mathcal A}}
\newcommand{\cF}{{\mathcal F}}
\newcommand{\cB}{{\mathcal B}} 
\newcommand{\cQ}{{\mathcal Q}} 
\newcommand{\cM}{{\mathcal M}} 
\newcommand{\cN}{{\mathcal N}} 
\newcommand{\llambda}{\lambda_B} 

\newcommand{\mmpair}{\mu^+\mu^-}
\renewcommand{\Re}{{\rm Re}}

\renewcommand{\Re}{\text{Re}}
\renewcommand{\Im}{\text{Im}}

\begin{document}

\title{\textbf{Disentangling short- vs.~long-distance dynamics in $B\to K^{*}\mu^+\mu^-$}}
\author[1]{Marzia Bordone\thanks{marzia.bordone@uni-mainz.de}}
\author[2]{Gino Isidori \thanks{isidori@physik.uzh.ch}}
\author[2]{Arianna Tinari \thanks{arianna.tinari@physik.uzh.ch}}

\affil[1]{\small{\textit{PRISMA}$^{++}$ Cluster of Excellence \& Mainz Institute for Theoretical Physics
Johannes Gutenberg University, Staudingerweg 9, D-55128 Mainz, Germany}}
\affil[2]{\small{Physik-Institut, Universit\"at Z\"urich, CH-8057 Z\"urich, Switzerland}}

\published{\flushright 
\footnotesize{MITP-26-032, ZU-TH 22/26}
\vskip2cm }

\maketitle

\begin{abstract}
We present an updated data-driven analysis of short- vs.~long-distance dynamics in $B\to K^{*}\mu^+\mu^-$. 
The analysis is based on the recent LHCb measurement of the full angular distribution of this process, taking advantage of the previously published global amplitude fit. Thanks to the precise di-lepton invariant-mass binning and angular observables directly sensitive to absorptive phases, current data provide novel insights into non-local contributions.  We analyze the latter using a dispersive description, whose free parameters are determined via the amplitude fit. The dispersive model provides a satisfactory overall description of current data, without the need to introduce additional {\em ad hoc} hadronic parameters.  The extracted value of the short-distance coefficient $C_9$ is lower than its Standard Model prediction, but significantly closer to it than in analyses where long-distance contributions are not constrained by data. The overall tension with the SM does not exceed the $2\sigma$ level.  We also obtain precise postdictions for the angular observables $S_{7,8,9}$, and envisage novel consistency tests among them. In view of future, more precise data, these could provide valuable tests of the description of non-local effects in this process that, if further validated, would enable an even more robust extraction of short-distance information.
\end{abstract}

\newpage 

\section{Introduction}

Flavor-changing neutral-current (FCNC) transitions of the type $b\to s\ell^+\ell^-$ are very interesting systems both to probe physics beyond the Standard Model (SM) and to better understand the SM itself. The strong suppression of these processes occurring within the SM implies an enhanced sensitivity to physics above the electroweak scale. At the same time, these modes provide a unique opportunity to deepen our understanding of the interplay between weak interactions and non-perturbative QCD. Extracting short-distance information from these transitions requires precise control of long-distance hadronic contributions, which remains the main theoretical challenge. As we shall discuss in this paper, in specific exclusive modes, the interplay between precise experimental data and theoretical methods provides a promising strategy to tackle this challenge. 

Among all exclusive $b\to s\ell^+\ell^-$ decays, the $B\to K^{*}\mu^+\mu^-$ mode plays a special role. 
On the one hand, the clean experimental signature allows precise measurements of this process at hadron colliders~\cite{CMS:2013mkz,LHCb:2013ghj,CMS:2015bcy,LHCb:2015svh,CMS:2017rzx,LHCb:2020lmf,CMS:2024atz,ATLAS:2018gqc}. 
On the other hand, the rich differential structure of the 
decay distribution makes it  particularly suitable  to investigate the interplay between short and long-distance 
dynamics~\cite{Altmannshofer:2008dz,Bobeth:2010wg,Descotes-Genon:2013vna}. The focus of this paper is the investigation of the vector-current part of the $B\to K^{*}\mu^+\mu^-$ decay amplitude, whose short-distance component is controlled by the Wilson coefficient $C_9$. A series of previous analyses has found the value of $C_9$ determined by data lies below its SM  prediction (see e.g.~\cite{Capdevila:2023yhq} and references therein). However, the vector-current part of the amplitude is the one receiving sizable non-local hadronic contributions. The main goal of this work is to quantify these long-distance effects using both recent data and theoretical constraints, and to assess their impact on the extraction of the short-distance coefficient $C_9$.

The purpose of this paper is twofold. First, we present an updated determination of $C_9$ employing the dispersive description of long-distance effect proposed in~\cite{Cornella:2020aoq,Bordone:2024hui}, taking advantage  
of new LHCb measurements based on the full Run-I and Run-II data sample. These include both a model-independent analysis of the full angular distribution~\cite{LHCb:2025mqb}, 
for different dilepton invariant mass ($q^2=m^2_{\mu\mu}$) bins, and a global amplitude fit~\cite{LHCb:2024onj} employing the dispersive framework. 
 Second, we exploit these new measurements to test the validity of the dispersive description itself and to assess its impact on the extraction of short-distance information.

In parallel with the new experimental analyses by LHCb~\cite{LHCb:2024onj,LHCb:2025mqb},
significant theoretical progress has been achieved recently in the study of long-distance effects in $B\to K^{(*)}\mu^+\mu^-$~\cite{Frezzotti:2025hif,Capdevila:2025drq,Hurth:2025vfx,
Mutke:2024tww,Hoferichter:2026jlh,Isidori:2024lng,Isidori:2025dkp,Altmannshofer:2026cwk,Ciuchini:2026hxy}.
Particularly noteworthy is the possibility, pointed out in~\cite{Frezzotti:2025hif},  
of estimating 
long-distance hadronic contributions in these processes via Lattice QCD. 
Although this possibility remains far from a realistic implementation, especially for the $K^*$ mode, it represents an important long-term perspective. Of more direct relevance to the present work are the recent studies of rescattering effects associated with anomalous thresholds~\cite{Mutke:2024tww,Hoferichter:2026jlh,Isidori:2024lng,Isidori:2025dkp}. On the one hand, the explicit estimates presented in Refs.~\cite{Isidori:2024lng,Isidori:2025dkp} for the $B\to K$ mode show that these contributions exhibit a non-trivial $q^2$ dependence, making them unlikely to mimic a universal short-distance contribution. On the other hand, Ref.~\cite{Hoferichter:2026jlh} demonstrated that such effects are compatible with the dispersive treatment and possess well-defined partonic counterparts, which can be estimated in perturbation theory in the kinematical regime where the operator-product expansion applies. These developments provide an important theoretical basis to support and interpret the present data-driven analysis via the dispersive description of long-distance effects. 

As first noted in Ref.~\cite{Altmannshofer:2026cwk}, the recent LHCb measurements of the $B\to K^{*}\mu^+\mu^-$ angular distribution provide clear evidence for non-local contributions already in the region $q^2\lesssim6~{\rm GeV}^2$, where these effects had often been assumed to be negligible. As discussed in Ref.~\cite{Altmannshofer:2026cwk}, this observation raises the possibility that long-distance effects could account for the long-standing tension in the determination of  $C_9$. Our framework allows this hypothesis to be tested quantitatively. The dispersive model {\em predicts} the long-distance contributions throughout the low-$q^2$ region in terms of parameters determined independently from data collected around the narrow charmonium resonances. The comparison with the new LHCb measurements therefore provides a genuine {\em validation} test of the dispersive framework, and this allows us to {\em quantify} the impact of long-distance dynamics on the extraction of $C_9$.

The improved sensitivity to rescattering effects arises from two main features of the new data. First, the finer $q^2$ binning reveals a much richer local structure of the decay amplitudes. Second, and more importantly, the angular observable $S_7$, which is directly proportional to the absorptive part of the decay amplitude, provides a particularly clean probe of non-local phases. In the region below the $J/\psi$ resonance, the observed rescattering effects are well described by our dispersive treatment, which is further based on the assumption 
that single-particle intermediate states (i.e.~vector mesons)
 saturate the dispersion relation in the $q^2$ variable. 
 It is important to stress that this constitutes an almost parameter-free prediction: the dominant resonance parameters are determined from data collected around the resonance peaks, which are excluded from the subsequent fit to the low-$q^2$ region. The latter is performed allowing independent values of $C_9$ in every $q^2$ bin and for each helicity amplitude, providing a stringent test of the dispersive parameterization. The extracted values of $C_9$ turn out to be compatible with an approximately helicity- and $q^2$-independent contribution, barring statistical fluctuations, as expected for short-distance dynamics. Current data therefore provide a non-trivial validation of the dispersive description while simultaneously allowing a more reliable extraction of the underlying short-distance contribution. As we shall show, the extracted value of   $C_9$ is lower than its SM prediction, but significantly closer to it than in most previous analyses: the overall tension with the SM does not exceed the $2\sigma$ level. 

A particularly interesting outcome of our analysis concerns the angular observables $S_7$, $S_8$, and $S_9$. Despite the sizable uncertainties affecting the overall magnitude of the decay amplitude, we obtain precise postdictions for these quantities thanks to the constrained structure of the absorptive part of the amplitude within the dispersive framework. These postdictions are significantly more precise than the corresponding current experimental measurements. We also show that $S_{7,8,9}$ satisfy a non-trivial sum rule that, to a good approximation, holds independently of the parameterization of rescattering effects. Future measurements of these observables can therefore provide valuable tests of the overall description of the decay amplitudes, as well as important internal consistency checks of the experimental data.

The paper is organized as follows. 
In Sec.~\ref{sect:theory}, we introduce the amplitude decomposition employed throughout the analysis. In Sec.~\ref{sect:THerror} we discuss the theoretical limitations of the  dispersive parameterization. Approximate expressions for $S_7$, $S_8$, and $S_9$, together with the corresponding sum rule, are presented in Sec.~\ref{sect:SiTH}. The results of the bin-by-bin fits to $C_9$, the comparison with the constant $C_9$ fit, and 
a general discussion of the results and our postdictions for $S_5$ and 
$S_{7,8,9}$ are presented in Sec.~\ref{sect:fit}. At the end of this section, a brief comparison of our determination of $C_9$ with those obtained in previous studies is also reported.  
The conclusions are summarized in Sec.~\ref{sect:Conc}.

\section{Amplitude decomposition}
\label{sect:theory}

\subsection{Effective Lagrangian and decay amplitude}
The effective Lagrangian describing  $b\to s\ell^+\ell^-$ transitions, after integrating 
out the SM degrees of freedom above the 
$b$-quark mass,
can be written as 
    \begin{equation}
       \cL_\mathrm{eff}(b\to s\ell^+\ell^-) = 
        \frac{4 G_F}{\sqrt{2}}  V_{tb}V^*_{ts}
  \sum_{i=1}^{10} C_i \cQ_i  ~ + ~\cL_{\rm QCD+QED}^{[N_f=5]} \,.
  \label{eq:Leff}
    \end{equation}
    Here $V_{ij}$ denote the elements of the Cabibbo-Kobayashi-Maskawa (CKM) matrix. To shorten the notation, the subleading terms proportional to  
    $V_{ub}V^*_{us}$ are not explicitly shown in Eq.~(\ref{eq:Leff}).
    The most  relevant effective operators are 
    \begin{align}
        \cQ_1 =& (\bar{s}^\alpha_{L}\gamma_\mu c^\beta_L)(\bar{c}^\beta_L\gamma^\mu b^\alpha_L)\,,   &  \cQ_2 =&(\bar{s}_{L}\gamma_\mu c_L)(\bar{c}_L\gamma^\mu b_L)\,,   \\
        \cQ_7=&\frac{e}{16\pi^2}m_b(\bar{s}_{L}\sigma^{\mu\nu} b_R)F_{\mu\nu}\,,    & \cQ_8=&\frac{g_s}{16\pi^2}m_b(\bar{s}_{L}\sigma^{\mu\nu}T^a b_R)G_{\mu\nu}^a\,,   \\
        \cQ_9=&\frac{e^2}{16\pi^2}(\bar{s}_{L}\gamma_\mu b_L)(\bar{\ell}\gamma^\mu \ell)\,,    &  \cQ_{10}=&\frac{e^2}{16\pi^2}(\bar{s}_{L}\gamma_\mu b_L)(\bar{\ell}\gamma^\mu\gamma_5 \ell)\,.
    \end{align}
    The explicit form of the additional four-quark operators
    $\cQ_{3-6}$, with 
    subleading Wilson coefficients, can be found in~\cite{Altmannshofer:2008dz}.
 
    Only the FCNC quark bilinears 
    $\cQ_{7,9,10}$ have 
    non-vanishing tree-level matrix elements in $B\to K^{(*)}\mmpair$.
    They yield the following amplitude 
\begin{eqnarray}
    &&  \left.\cM\left(B\rightarrow K^*\ell^+ \ell^- \right)\right\vert_{C_{7,9,10}} 
        = \cN\,   \Bigg\{  \no \\
    &&  \quad    
            - \left[ C_L + 
            \frac{ 2 m_b (m_B+m_{K^*} )}{ q^2} 
            \frac{ T_1(q^2)}{V(q^2)} C_7 \right]
            \frac{   2 V(q^2) }{m_B+m_{K^*}} 
            i \epsilon_{\mu\nu\rho\sigma} (\epsilon^{\ast})^\nu p_B^\rho p_{K^*}^\sigma  
            \no\\
    &&      \quad -\left[ C_L
            + \frac{2 m_b (m_B+m_{K^*})}{q^2} \frac{T_2(q^2)}{A_2(q^2)} C_7 \left(1+ O\Big(\frac{q^2}{m_B^2}\Big)\right) \right]
            \frac{  A_2(q^2) }{ m_B+m_{K^*} }     (\epsilon^{\ast} \cdot q)(p_B+p_{K^*})_\mu 
            \no\\
    &&      \quad +\left[C_L
            + \frac{ 2 m_b(m_B^2- m_{K^*})}{q^2} \frac{T_2(q^2)}{A_1(q^2)} C_7 \right] A_1(q^2) \left(m_B + m_{K^*} \right) \epsilon_{\mu}^{\ast} 
            \Bigg\}\,  J^\mu_{\ell,L}  + (L\to R)\,,   
            \quad
            \label{eq:C9matrixel}    
\end{eqnarray}
where 
\begin{equation}
 J^\mu_{\ell,L} = \bar \ell_L \gamma^\mu \ell_L \,,  \qquad 
 J^\mu_{\ell,R} = \bar \ell_R \gamma^\mu \ell_R \,,
\qquad  C_{L,R} = C_9 \mp C_{10}\,,
 \end{equation}
 and 
 \begin{equation}
 q^\mu=p_B^\mu-p_{K^{(*)}}^\mu\,, \qquad 
\cN=  \sqrt{2} G_\mathrm{F} \alpha_{\rm em} V_{tb} V_{ts}^\ast/(4\pi)\,, 
\end{equation}
while  
$\{A_i(q^2), V(q^2), T_i(q^2)\}$ 
 denote the 
$B\to K^*$ hadronic local form factors. 
The independent Lorentz structures
appearing in these amplitudes are in a linear relation with the 
three independent  $|B\rangle \to |K^*_\lambda \rangle$ helicity 
amplitudes ($\lambda=\perp,\parallel,0$). For later purposes, 
following the notation of Ref.~\cite{Bordone:2024hui}, we define
the following form-factor combinations  
\begin{align}
  \cF_\perp(q^2) &= V(q^2) \,, \nonumber \\ 
 \cF_\parallel(q^2) &=  A_1(q^2) \,,  
  \nonumber \\ 
  \cF_0(q^2)  &=  \frac{ (m_B+m_{K^*})^2( m_B^2-m_{K^*}^2-q^2) A_1(q^2)
- \lambda_B(q^2) A_2(q^2)}{16 m_B m_{K^*}^2 (m_B+m_{K^*})}\,,
\label{eq:helicities}
\end{align}
with $\lambda_B(q^2)= (m_B^2-m_{K^*}^2-q^2)^2 - 4 q^2 m_{K^*}^2$. 
Normalising the helicity amplitudes, $\cA^{L,R}_\lambda(q^2)$, such that, after integrating over all the angular variables,  
\begin{align}
\frac{ d\Gamma (B\to K^*\ell^+\ell^-)}{dq^2}  = \sum_{\lambda = \perp,\parallel, 0} \Big[  |\cA^L_{\lambda}(q^2)|^2 + |\cA^R_{\lambda}(q^2)|^2 \Big]\,,
\label{eq:Al-norm}
\end{align}
the contribution proportional to  $C_{9,10}$ to the helicity amplitudes assumes the  form 
\begin{align}
\left. \cA^{L,R}_{\lambda} (q^2) \right|_{C_{9,10}} =  \kappa_\lambda(q^2)  \,  C_{L,R}  \, \cF_\lambda(q^2)\,,  \label{eq:kappal-def}
\end{align}
with the $\kappa_\lambda(q^2)$ reported in Appendix~\ref{app:etaV}.

\medskip 

The theoretical challenge in estimating the theoretical amplitude lies in the non-local matrix elements of the four-quark operators $\cQ_{1-6}$, which encode rescattering effects. 
To all orders in $\alpha_s$, and to first order in $\alpha_{\rm em}$, these have the same structure as the matrix elements of $\cQ_7$ and $\cQ_9$. In other words, Lorentz and gauge invariance imply~\cite{Bordone:2024hui}  
\begin{align}
 & \left.\cM\left(B\rightarrow K^*_\lambda\, \ell^+ \ell^-\right)\right\vert_{C_{1-6}}  =\ - i \frac{ 32 \pi^2\cN}{q^2}\,  \bar \ell \gamma^\mu \ell  
\int d^4 x e^{iqx} \langle K^*_\lambda   | T\left\{     
j_{\rm \mu}^{\rm em}(x), \sum_{i=1,6} C_i\cQ_i (0)
\right\} | B  \rangle \nonumber \\
& \qquad\qquad  = 
 \left( \Delta^\lambda_9(q^2) + \frac{m_B^2}{q^2} \Delta^\lambda_7
\right) \langle K^*_\lambda\ \ell^+ \ell^- |  \cQ_9  | B  \rangle \,,
\label{eq:MQi}
\end{align}
where  $j_\mu^\text{em} (x)$ denotes the electromagnetic current 
and the explicit form of $\langle K^*_\lambda\, \ell^+ \ell^- |  \cQ_9  | B  \rangle$ follows from Eq.~(\ref{eq:C9matrixel}). 
Here $\Delta^\lambda_9(q^2)$ is a function regular in the limit $ 
q^2\to0$, while $\Delta^\lambda_7$  is 
$q^2$ independent and, by helicity conservation, is non-vanishing only for  $\lambda=\perp$ and $\parallel$.

Thanks to the decomposition in Eq.~(\ref{eq:MQi}), we can effectively describe these matrix elements via 
an appropriate $q^2$-- and $\lambda$--dependent modification of  $C_9$ and a  $q^2$--independent shift of $C_7$ in Eq.~(\ref{eq:C9matrixel}).

\subsection{Dispersive description of the non-local matrix elements}
\label{sect:nonpert4q}

The non-local matrix elements in Eq.~(\ref{eq:MQi}) are non-perturbative quantities that cannot be estimated in perturbative QCD for generic $q^2$ values. Following a procedure outlined in~\cite{Khodjamirian:2012rm,Lyon:2014hpa,Blake:2017fyh,Bobeth:2017vxj,Cornella:2020aoq}, we estimate them using a subtracted dispersion relation, combined with a subtraction point in a kinematical region where the $T$-ordered product in Eq.~(\ref{eq:MQi}) can be estimated reliably in perturbation theory. 

More precisely, we write a one-time subtracted dispersion relation for the three hadronic functions
$\cH^\lambda(q^{2})$, defined by 
\begin{align}
  &  - 2i \int d^4x e^{iqx}\langle K^* |T\left\{j_\mu^\text{em}(x), \sum_{i=1,6} C_i \cQ_i (0) \right\}| B\rangle = 
        \nonumber \\  
&   \quad  =  \left(\epsilon_{\mu}^{\ast} 
            -q_\mu \frac{\epsilon^{\ast}\cdot q }{q^2}\right) 
            \left(m_B + m_{K^*} \right) \cH^{\parallel}(q^2)
            -i \epsilon_{\mu\nu\rho\sigma} (\epsilon^{\ast})^\nu p_B^\rho p_{K^*}^\sigma 
            \frac{2 }{m_B+m_{K^*}} \cH^\perp(q^2)  
            \nonumber \\
     &\quad    -   \left(\left(p_B+p_{K^*} \right)_\mu -q_\mu \frac{q\cdot (p_B+p_{K^*})}{q^2} \right) 
            \frac{\epsilon^{\ast} \cdot q}{ m_B+m_{K^*} }  \tcH^{0}(q^2)\,, 
            \label{eq:BKsdec}
\end{align}
with
\begin{equation}
      \cH^0(q^{2})  =
        \frac{(m_B+m_{K^*})^2 ( m_B^2-m_{K^*}^2-q^2) \cH^\parallel(q^{2})
        -\lambda(m_B^2,m_{K^*}^2, q^2)   \tcH^0(q^{2})}{16m_B m_{K^*}^2 (m_B+m_{K^*})}\,.  
\end{equation}
Note that, contrary to Ref.~\cite{Bordone:2024hui},
we do not limit ourselves only to the leading 
$\cQ_{1,2}$ operators, and $c\bar c$ intermediate states, 
but we also keep four-quark operators with light quarks.

The subtracted dispersion relation for each $\cH^\lambda(q^{2})$  reads  
        \begin{align}
        \begin{aligned}
            \Delta \cH^\lambda(q^{2})  = \frac{q^2- q_{0}^2}{2\pi i} \int_{s_0}^\infty ds \frac{ {\rm Disc}[\cH^\lambda(s)]}{(s-q_{0}^{2})(s-q^2)}  \equiv \frac{q^2-q_{0}^2}{\pi}\int_{s_0}^\infty ds \frac{\rho^\lambda(s)}{(s-q_{0}^{2})(s-q^2)}\,.
        \label{eq:DH_gen}    
        \end{aligned}
        \end{align}   
Using this result, we can express the terms in Eq.~(\ref{eq:MQi}) which are regular in the $q^2\to 0$ limit 
via the substitution 
\begin{equation}
    C_9 \to C_9^{\mathrm{eff}, \lambda} (q^2) = C_9 + Y^\lambda (q^2)~,
    \label{eq:C9eff}
\end{equation}
where  
 \begin{align}
Y^\lambda(q^{2}) =  Y^\lambda (q^2_0) +  \frac{16 \pi^{2}}{\cF_\lambda (q^2)} \Delta \cH^\lambda(q^{2})\,,
\label{eq:YdecNP}
 \end{align}           
with the $\cF_\lambda (q^2)$ defined in Eq.~(\ref{eq:helicities}). The only hypothesis under which this decomposition holds is the analyticity of $\cH^\lambda(q^{2})$ with respect to $q^{2}$, but for the discontinuity on the real axis for $q^{2} \geq s_0 \equiv 4 m_\pi^2$.\footnote{Note that $\cH^\lambda(q^{2})$ is a complex function also for $q^{2} \leq s_0$, due to absorptive contributions not related the cut in $q^{2}$. }

To make concrete predictions, we employ two further, stronger hypotheses (with two related sets of approximations): 
\begin{enumerate}
    \item[i.] $Y^\lambda (q^2_0)$ can be computed reliably in perturbation theory for  
    $q^2_0 =-({\rm few~GeV})^2$;
    \item[ii.] the spectral function $\rho^\lambda(s)$ in (\ref{eq:DH_gen})
    is dominated by the contributions of single-particle intermediate states, namely by the contribution of vector-meson resonances.
\end{enumerate}
According to the second hypothesis, we decompose $\Delta \cH^\lambda(q^{2})$
as
\begin{align} 
            \Delta \cH (q^2) = \sum_{V}\eta^\lambda_V e^{i\delta^\lambda_V}\frac{(q^2-q_0^2)}{(m_V^2-q_0^2)} A_V^\text{res}(q^2)\,,
          \label{eq:etaVdef}     
\end{align}
where 
\begin{align}
             A_V^\text{res}(q^2)  = \frac{m_V \Gamma_V}{m_V^2-q^2-i m_V \Gamma_V}\,.
\end{align}
The sum in (\ref{eq:etaVdef}) extends over all spin-1 states $V$ for which the transition $B\to K^* V(\to \ell^+\ell^-)$ is allowed. Following the data analysis presented in Ref.~\cite{LHCb:2024onj}, we limit this sum to five charmonium resonances and the three lightest vector resonances ($\rho,\omega,\phi$). The coefficients $\eta^\lambda_V$ and $\delta^\lambda_V$, which are determined by data, are reported in Appendix~\ref{app:etaV}. 

\paragraph{Subtraction point.}
For negative $q^2$ values, far from any physical threshold with respect to the dilepton invariant mass, 
it is a good approximation to evaluate the $T$-product in Eq.~(\ref{eq:MQi}) in perturbation theory 
in the heavy-quark limit. Neglecting power corrections in the heavy-quark masses, the result is helicity independent. Using the next-to-leading (NLO) analysis in~\cite{Asatrian:2019kbk} one gets
\begin{equation}
    Y_{\rm pert}(q_0^2)_{\rm LO+NLO} = 0.13 \times (1-0.02i)\,, 
    \qquad q_0^2= -4.6~{\rm GeV}^2\,,
    \label{eq:Yq0}
\end{equation}
to be compared with the  $O(\alpha_s^0)$ result 
$Y_{\rm pert}(q_0^2)_{\rm LO} = 0.23$. The numerical value in 
Eq.~(\ref{eq:Yq0}) has been obtained setting $m_c/m_b=0.29$ 
in the $\overline{\rm MS}$ scheme) and fixing  the renormalization scale to $\mu_b = 4.2~{\rm GeV}$. 
The choice $q_0^2= -4.6~{\rm GeV}^2$ for the 
subtraction point is motivated by the analog choice made in the experimental analysis~\cite{LHCb:2024onj}, where most of the parameter couplings are extracted. In principle, the complete result for $Y^\lambda(q^2)$
 should not depend on the choice of $q_0^2$. However, a residual dependence appears, given the approximations employed.
We comment on this aspect, as well as the other uncertainties related to the subtraction point, in Section~\ref{sect:THerror}.  

\paragraph*{Non-local matrix-elements singular in the $q^2\to 0$ limit.}
Concerning the  $\Delta_7^\lambda$ terms in Eq.~(\ref{eq:MQi}), on general grounds we can write 
\begin{align}
C_7 ~\to~ &{\rm Re}(C_7^{\mathrm{eff},\lambda}) + i\, 
{\rm Im}(C_7^{\mathrm{eff},\lambda}) \qquad  (\lambda=\perp,\parallel)~\,.
\label{eq:C7eff1}
\end{align}
Since the result is expected to be helicity-independent in the heavy-quark limit, in order to reduce the number of free parameters in the following we assume 
$C_7^{\lambda,\mathrm{eff}}  \approx C_7^{\mathrm{eff}}$.

Using the results of Ref.~\cite{Beneke:2001at}, where the four-quark matrix elements have been estimated to NLO accuracy in $\alpha_s$, we can write 
\begin{equation}
\qquad {\rm Re} (C_7^\mathrm{eff}) \approx 1.33\, C_7\,.
\label{eq:C7eff}
\end{equation}
We implement this shift in 
all $C_7$ terms in 
Eq.~(\ref{eq:C9matrixel}). 
In order to take into account the scale uncertainty,  missing higher-order corrections,
and power corrections, 
we assign a conservative $10\%$ error
to the numerical result in 
Eq.~(\ref{eq:C7eff}).

Higher-order and power corrections are expected to also generate a non-vanishing 
${\rm Im} (C_7^{\rm eff})$ 
of $O(10^{-2})$~\cite{Khodjamirian:2010vf,Altmannshofer:2014rta}. In Ref.~\cite{Paul:2016urs} the explicit estimate ${\rm Im} (C_7^{\rm eff})=-0.027\pm0.016$ has been obtained. 
This absorptive term has negligible impact in most differential distributions; however, it plays an important role in the low-$q^2$ behavior of $S_{7,8,9}$ hence cannot be neglected (see Sect~\ref{sect:SiTH}).
As discussed in Sect~\ref{sect:C9fit} we determine the range of  ${\rm Im} (C_7^{\rm eff})$ combining both theoretical constraints and current data.

\section{Theoretical uncertainties of the dispersive approach}
\label{sect:THerror}

\subsection{The subtraction point}
In principle, the functions $\cH^\lambda(q^2)$
encoding the $B\to K^* \ell^+\ell^-$ 
non-local matrix elements of four-quark operators do not depend on the subtraction point. However, 
the decomposition in terms of resonance and 
perturbative terms, via the choice of a specific subtraction point,
introduces an unphysical dependence on $q^2_0$.
The latter reflects the  uncertainty
due non-perturbative contributions to $\cH^\lambda(q^2)$ that we are not able to estimate precisely for any choice of the subtraction point. 
 
Explicit estimates and general arguments suggest that non-perturbative contributions to $\cH^\lambda(q^2)$  are small both at $q^2 \lesssim 0$ and for very large positive and negative  $q^2$ values  (i.e.~far from the region with on-shell narrow-resonance terms). In the following, we investigate the parametric structure of results obtained with different, motivated, subtraction points in order to estimate the uncertainty associated with a specific choice of $q_0^2$. 
  
The uncertainty on the subtraction point 
translates into a constant ($q^2$-independent) 
helicity-dependent shift in $C_9$. We can estimate this effect by comparing the value of $\cH^\lambda (q^2=0)$  for two extreme, well-motivated choices of $q_0$, namely $q^2_0=0$ and $q^2_0 \to \pm \infty$. By doing so, we obtain 
\begin{equation}
\Delta C^\lambda_{9,\, {\rm sub}} = \frac{16 \pi^2}{\mathcal{F}^\lambda(0)} \Bigg(
\Delta \cH^\lambda (0)\Big|_{q_0 \to \pm \infty }
- \Delta \cH^\lambda (0)\Big|_{q_0 \to 0 } \Bigg)
=  \frac{16\pi^2}{\cF^\lambda(0)}
 \sum_{V } \eta^\lambda_V e^{i ( \delta^\lambda_V 
 +\phi_V)} 
   \frac{ \Gamma_V }{ \sqrt{m^2_{V} + \Gamma^2_V} } 
   \,,
\end{equation}
where $\phi_V = \arctan( \Gamma_V/m_V)$. Given the  dependence on masses, form factors, and coupling constants contributing to 
$\eta^\lambda_V$ and $\Gamma_V$, we deduce the following general parametric structure for the contributions to $\Delta C^\lambda_{9,\, {\rm sub}}$
by the leading resonances\footnote{The appearance of $\Lambda_{\rm QCD}$ is a consequence 
of expressing partial decay widths in terms of 
masses (and momenta),  couplings, and form factors.
We use in particular $\Gamma(V \to \ell^+\ell^-) \sim \alpha^2 \Lambda_{\rm QCD}^2 /m_V$ for both heavy and light resonances.}
\begin{equation}
| \Delta C^\lambda_{9,\, {\rm sub}} |_V  \, \approx  \,
\pi^2 \frac{\Lambda^2_{\rm QCD} m_B}{m_V^3} 
\times \left\{ \begin{array}{ll} O(1) & V = J/\Psi, \Psi (2S)\,,  \\[5pt]
O(|C_{3\ldots6}/C_2|)\quad & V = \rho,\omega,\phi\,. \\
\end{array}
\right.
 \end{equation}
By looking at this expression, we deduce the following two conclusions:
\begin{itemize}
\item {\bf Charm resonances}. In this case, the effect is small, of $O(0.1)$, as confirmed by the explicit numerical evaluation given the experimental values of the $\eta^\lambda_V$. 
The induced contribution is a genuine non-perturbative effect scaling as $\Lambda^2_{\rm QCD} m_b /m_c^3$,
which is independent of the perturbative contribution of $O(\alpha_s^n)$.
\item {\bf Light resonances}.
Here, the parametric dependence of individual resonances is not infrared safe. The numerical impact of this fact is not very large, given the smallness of the corresponding Wilson coefficients (and the CKM suppression of the $b\to s u\bar u$ amplitudes); however,  this signals a potential inconsistency in the parameterization of light-resonance contributions.
The infrared sensitivity cancels when all light resonances are included, if the following sum rule 
\begin{equation}
\sum_{V=\rho,\omega,\phi} \eta^\lambda_V e^{i ( \delta^\lambda_V 
 +\phi_V)} 
   \frac{ \Gamma_V }{ \sqrt{m^2_{V} + \Gamma^2_V} } = 0
\end{equation}
holds. Using the numerical results of the $\eta^\lambda_V$ and  $\delta_V^\lambda$
reported in Appendix~\ref{app:etaV},
we have checked that the sum rule is fulfilled by the current data (within errors). This provides a non-trivial a posteriori consistency check of the approach. In view of future data analyses, it would be useful to impose the sum rule to reduce uncertainties. 
\end{itemize}

Besides the above considerations on the resonance contributions, an additional uncertainty in the evaluation of the subtraction term stems from the uncertainty of the perturbative result in Eq.~(\ref{eq:Yq0}). In the following, we employ the following numerical 
values
\begin{equation}
    Y(q_0^2) 
    = 0.13\,,
    \qquad 
    \Delta Y(q_0^2)_{\rm th} 
    = \pm 0.21\,, \qquad  
    \label{eq:Yq0-fin}
\end{equation}
for $q_0^2=-4.6~{\rm GeV}^2$ (we drop the small imaginary part in Eq.~(\ref{eq:Yq0}) which has a negligible impact).
The estimate of $\Delta Y(q_0^2)_{\rm th}$ is obtained combining in quadrature: 
\begin{enumerate}
\item[i.]
 the uncertainty resulting from the variation of the renormalization scale ($\pm 50\%$ around $\mu_b=4.2$~GeV) on the perturbative term, which amounts to 
  $\pm 0.15$;
 \item[ii.] the uncertainty estimated by the explicit variation of the subtraction 
point (between 0 and $-4.6~{\rm GeV}^2$) on the resonance terms,
which amounts to $\pm 0.15$.
\end{enumerate}

\subsection{Explicit results for the non-perturbative contribution}
\label{sect:C9_LD}

\begin{figure}[t]
    \centering
\includegraphics[width=0.80\linewidth]{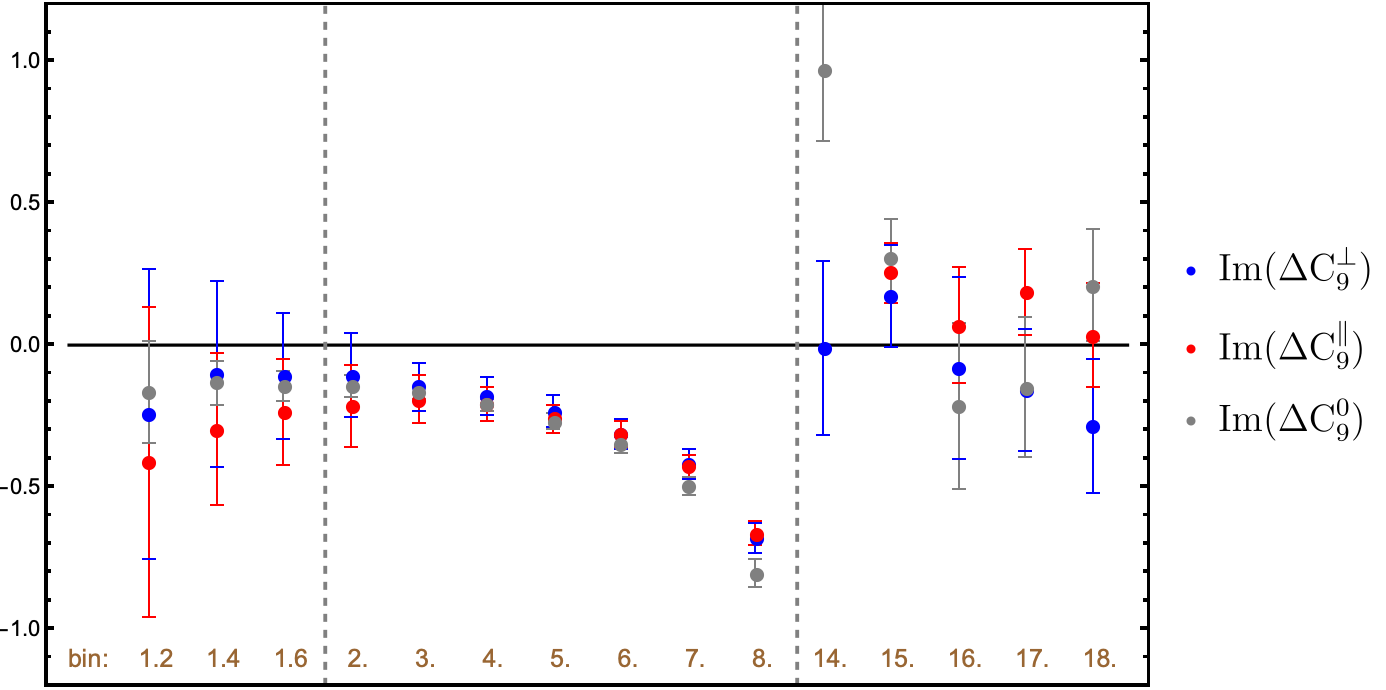}\\[8pt]
\includegraphics[width=0.80\linewidth]{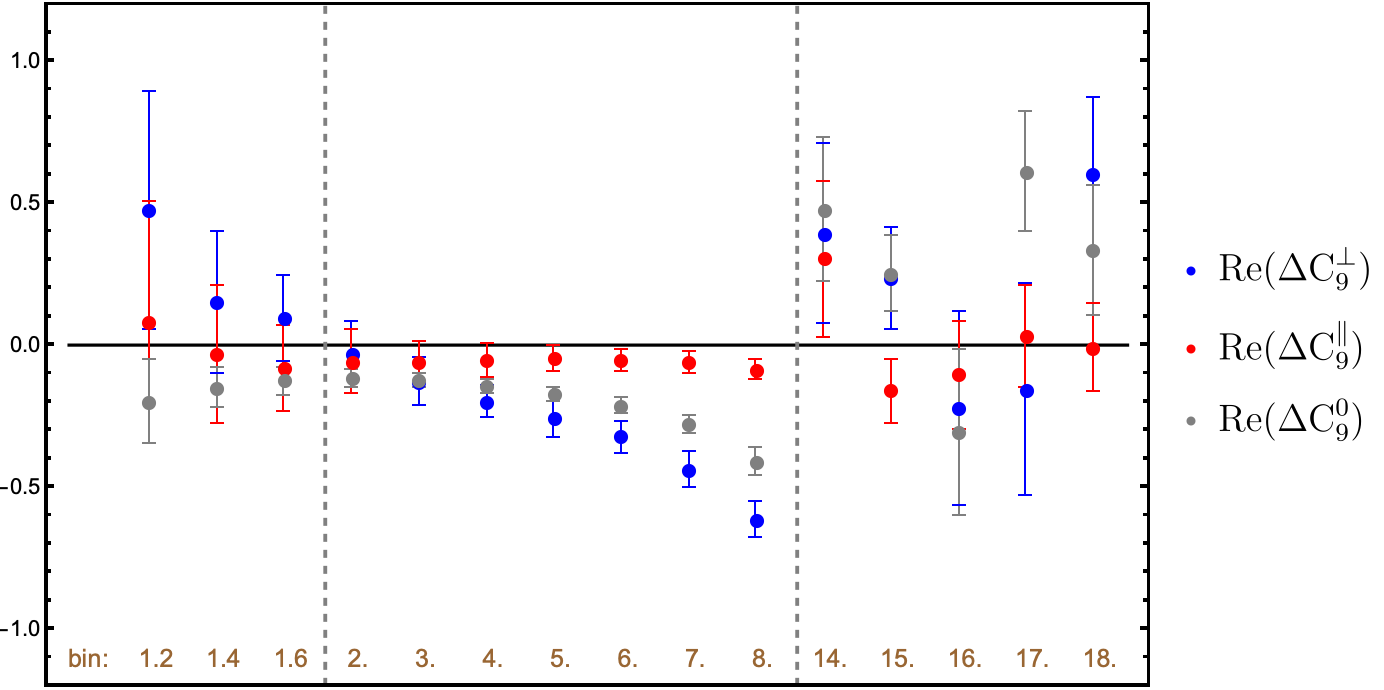}
    \caption{Long-distance contributions to the imaginary parts (upper plot)
    and the real part (lower plot) of $C^\lambda_9(q^2)$
    in different $q^2$ bins, as estimated by saturating the dispersion relation via the tower of 
    vector-meson resonances. 
    The gray lines separate three different $q^2$ regions. The $q^2$ bins, whose values in ${\rm GeV}^2$ are shown on the horizontal axis,  are equally spaced only within each region. 
    \label{fig:C9LD}}
\end{figure}

In Fig.~\ref{fig:C9LD} we show numerical results for the long-distance contribution to $C_9^\lambda(q^{2})$, namely
 \begin{align}
 \Delta C_9^\lambda(q^{2})_{\rm res} =  \frac{16 \pi^{2}}{\cF_\lambda (q^2)} \sum_{V}\eta^\lambda_V e^{i\delta^\lambda_V}\frac{(q^2-q_0^2)}{(m_V^2-q_0^2)} A_V^\text{res}(q^2)\,,
 \label{eq:C9LD}
 \end{align}  
 using the experimental results for the resonance parameters in Appendix~\ref{app:etaV}. We show both real and imaginary parts of $\Delta C_9^\lambda(q^{2})_{\rm res}$
for different (discrete) $q^2$ values. The error bars shown in the plots indicate the $68\%$~CL range resulting from the uncertainty on the resonance parameters (the local form factors are fixed to their central values). 

As can be seen from these plots, we can distinguish three main kinematical regions with different behaviors for $\Delta C_9^\lambda(q^{2})_{\rm res}$:
\begin{itemize}
\item {\bf Very-low $q^2$ region}  [$q^2 \lesssim 1.5~{\rm GeV}^2$]. 
Here, long-distance contributions are dominated by the light resonances and are rapidly varying in $q^2$, given the nearby poles of the light resonances. 
\item {\bf Low-$q^2$ region} [$1.5~{\rm GeV}^2 \lesssim q^2 \lesssim 8~{\rm GeV}^2$]. 
Here, long-distance contributions are dominated by the tails of the two narrow charmonium states. Since $q^2$ is sufficiently far from the pole of any vector state, $\Delta C_9^\lambda(q^{2})_{\rm res}$ is slowly varying in $q^2$. The absolute magnitude of $C_9^\lambda(q^{2})_{\rm res}$ is particularly small, 
$|C_9^\lambda(q^{2})_{\rm res}|\lesssim 0.3$, for $2~{\rm GeV}^2 \lesssim q^2 \lesssim 6~{\rm GeV}^2$. 
\item {\bf High-$q^2$  region}
[$q^2\gtrsim 14~{\rm GeV}^2$]. Here 
also the broader charmonium resonances play an important role, and long-distance contributions are both sizable and rapidly varying in~$q^2$.
\end{itemize}
We do not show the results in the narrow-resonance region explicitly, for $8~{\rm GeV}^2 \lesssim q^2 \lesssim 14~{\rm GeV}^2$,  where the long-distance contributions become comparable and even exceed the short-distance ones.
%
As we shall discuss in Section~\ref{sect:fit}, the behavior in all three $q^2$ regions shown in Fig~\ref{fig:C9LD} is important to interpret the results of the fit. 

\section{Approximate expressions for $S_{7,8,9}$}
\label{sect:SiTH}

The CP-averaged differential distributions $S_{7,8,9}$, defined as in~\cite{Altmannshofer:2008dz}, are particularly interesting, being directly proportional to the absorptive parts of the amplitudes, hence to rescattering effects. 
Encoding the latter via the modified 
coefficients $C^{{\rm eff},\lambda}_9(q^2)$ and 
$C^{{\rm eff},\lambda}_7$,
as discussed in Section~\ref{sect:nonpert4q},
leads to the following 
approximate theoretical expressions for 
these observables:\footnote{
We adopt here the notation of LHCb where $S_7$ and $S_8$ are defined with an opposite sign compared to Ref.~\cite{Altmannshofer:2008dz}.
To ease the notation, in the rest of this section we omit the superscript ``eff" in $C^{{\rm eff},\lambda}_9(q^2)$ and 
$C^{{\rm eff},\lambda}_7$.}
\begin{align}
S_7(q^2) &\approx  \kappa_7(q^2) \frac{ \Re(C_{10}) }{|C_{10}|^2 +|C_9(q^2)|^2 } \Bigg\{ \Im\Big[ C_9^\parallel(q^2) + \frac{m^2_B }{q^2}  \bar{C}^\parallel_7 \Big] - \Im[ 
C_9^0(q^2)]  \Bigg\} \,, 
\nonumber \\[3pt]
S_8 (q^2) &\approx  \kappa_8(q^2)  \frac{ \Re[C_{9}(q^2)] }{|C_{10}|^2 +|C_9(q^2)|^2 } \Bigg\{
\Im \Big[  C_9^\perp(q^2) + \frac{m^2_B }{q^2} \bar C^\perp_7\Big]- \Im[C_9^0(q^2)] \left[ 1 +\frac{2m_b m_B }{q^2}  \frac{\Re(C_7)}{\Re(C_9)} \right]  \Bigg\}\,, 
\nonumber \\[3pt]
 S_9 (q^2) &\approx  \kappa_9(q^2) \frac{ \Re[C_{9}(q^2)] }{|C_{10}|^2  +|C_9(q^2)|^2 } 
 \Bigg\{
\Im\Big[   C_9^\parallel(q^2) + \frac{m^2_B }{q^2} \bar C^\parallel_7
\Big]- \Im\Big[C_9^\perp(q^2) +
\frac{m^2_B }{q^2} \bar C^\perp_7  \Big]\Bigg\}\,, 
\label{eq:Sapp79}
\end{align}
where $\bar C^\lambda_7= 2 m_bC^\lambda_7 /m_B$,
\begin{align}
\kappa_7(q^2) &= - \frac{\sqrt{q^2} (m_B+m_{K^*})}{4 m_B m_{K^*}}  \frac{ \cF_\parallel(q^2) }{\cF_0(q^2) \cR(q^2) }\,,
\nonumber \\
\kappa_8(q^2) &= \frac{\sqrt{q^2 \lambda_B(q^2)} }{8 m_{K^*} m_B (m_B+m_{K^*})} 
 \frac{\cF_\perp(q^2) }{\cF_0(q^2) \cR(q^2)}   \,,  \nonumber \\
 \kappa_9(q^2) &= - \frac{  q^2 \sqrt{\lambda_B(q^2)} }{32 m^2_{K^*} m_B^2}
\frac{\cF_\perp(q^2) \cF_\parallel(q^2) }{\cF^2_0(q^2) \cR(q^2)}
 \left[ 1 +\frac{2m_b m_B }{q^2}  \frac{\Re(C_7)}{\Re(C_9)} \right]   \,, 
\end{align}
and
\begin{equation}
\cR(q^2) = 1 + \frac{q^2}{32 m_{K^*}^2 } \left[  \frac{\lambda_B(q^2)} {m_B^2 (m_B+m_{K^*})^2} \frac{ \cF^2_\perp(q^2)}{\cF^2_0(q^2)} +  
 \frac{(m_B+m_{K^*})^2} {m_B^2}  \frac{ \cF^2_\parallel(q^2)}{\cF^2_0(q^2)}   \right]\,.
\end{equation}
To obtain these expressions, we have neglected: i)~lepton masses; 
ii)~subleading terms induced by the dipole operator (whose leading contribution is evaluated in the large-recoil limit); 
iii)~helicity-dependent long-distance corrections in the real parts of the amplitudes, i.e.~we have set $\Re[C^\lambda_9(q^2)] \approx\Re[C_9(q^2)]$. These approximations hold to a good accuracy in most of the spectrum, with the exception of the singularity regions, namely the very-low $q^2$ region and that of the narrow-resonance peaks of $J/\Psi$ and $\Psi(2S)$.\footnote{
For $q^2$ close to zero, we cannot neglect lepton masses, and the dipole operator has a prominent role; very close to the narrow-resonance peaks $\Re(C^\lambda_9)$ is completely dominated by rescattering terms, hence the approximation 
$\Re(C^\lambda_9)\approx\Re(C_9)$ is no longer valid.}
This expectation is confirmed a posteriori by the extraction of $\Re[C^\lambda_9(q^2)]$, especially in the low-$q^2$ region, as defined in Sect.~\ref{sect:C9_LD}.

By means of Eq.~(\ref{eq:Sapp79}), determining independently $\Re(C_9)$ and $\Re(C_{10})$ leads to an approximate sum rule among the three $S_i$:\footnote{Beside the approximations already employed in Eq.~(\ref{eq:Sapp79}), in order to arrive to Eq.~(\ref{eq:SumRuleSi}) we have neglected the ${\rm Re}(C_7)/{\rm Re}(C_9)$ term in the numerator of $S_8$.}
\begin{equation}
\frac{ S_7(q^2) } { \kappa_7(q^2) \Re(C_{10})} 
- \frac{ S_8(q^2) } { \kappa_8(q^2) \Re[C_9(q^2)] } 
- \frac{ S_9(q^2) } { \kappa_9(q^2) \Re[C_9(q^2)] } \approx 0\,.
\label{eq:SumRuleSi}
\end{equation}
This sum rule holds regardless of the values of the absorptive parts of the amplitudes, hence it provides an important consistency check of the data with the SM hypothesis, irrespective of rescattering effects. 
More precisely, Eq.~(\ref{eq:SumRuleSi}) holds both in the SM and in any extensions of the SM with the same basis of effective operators. 

Two further points relevant for the theory-data comparison follow from the analytical expressions in Eq.~(\ref{eq:Sapp79}):
\begin{itemize}
\item
The contributions to $S_{7,8,9}$ generated by 
$\Im(C^\lambda_9)\not=0$ vanish for $q^2 \to 0$.
Hence, non-vanishing $S_{7,8,9}$ at very low $q^2$ values are dominated by absorptive terms in the dipole amplitude.
For instance, in the case of $S_7$ we get 
\begin{equation}
 S_7 \Big|_{q^2=1.5\, {\rm GeV}^2}
~\approx 1.8 \times 
\Im\big( C^\parallel_7\big) + 0.05 \times
\Im\big(C_9^\parallel-C_9^\perp\big) 
\Big|_{q^2=1.5\, {\rm GeV}^2}~.
\label{eq:S7lowq}
\end{equation}
This implies that if ${\rm Im}(C_9^\lambda) = O(10^{-1})$, as shown in Fig.~\ref{fig:C9LD},
the corresponding contribution
to $S_7$, for $q^2\approx 1.5\, {\rm GeV}^2$, is negligible. 
On the other hand, 
Eq.~(\ref{eq:S7lowq}) also implies that we cannot neglect 
${\rm Im}(C^\lambda_7)$, even if it is $O(10^{-2})$, when estimating $S_7$ at low $q^2$.
\item
In the low-$q^2$ region  the $\kappa_i$ satisfy 
$\kappa_7 > \kappa_8 \gg \kappa_9 $. Hence if the 
$\Im(C^\lambda_9)$ are of similar size, 
we should expect $|S_7| > |S_8| \gg |S_9|$.
\end{itemize}

It is instructive to compare our approximate expressions with those of Altmannshofer {\em et al.}~\cite{Altmannshofer:2026cwk}.
 To this purpose, we need to express our modified $C_i^\lambda$
 in terms of the $a_i$ and $b_i$ parameters of the polynomial expansion
 employed in~\cite{Altmannshofer:2026cwk}.
The following relations hold:
\begin{align}
   \delta \bar C_7^\perp &= a_- + a_+\,,
    \qquad\quad~ \delta \bar C_7^\parallel = a_- - a_+\,,
    \nonumber \\
    \delta C_9^0 &= a_0 + b_0 \frac{q^2}{m_B^2}\,, \qquad
    \delta C_9^\perp = b_- + b_+\,, 
    \qquad \delta C_9^\parallel = b_- - b_+\,.
    \label{eq:C9-truncated}
\end{align}
Using these expressions in Eq.~(\ref{eq:Sapp79}), and expanding to the relevant order, we recover the results in~\cite{Altmannshofer:2026cwk}. This comparison allows us to check how well the truncated expressions in Eq.~(\ref{eq:C9-truncated}) describe the phases determined by the resonance contributions. By looking at the results shown in Fig.~\ref{fig:C9LD} we deduce that:
\begin{itemize}
    \item In the region 
    $2~{\rm GeV}^2 \lesssim q^2 \lesssim 5~{\rm GeV}^2$,
    a linear expansion for the $C_9^\lambda$  is a reasonable approximation; however, to describe  a wider range in $q^2$
    more terms in the polynomial expansion are necessary. 
    \item Truncating the polynomials at different orders for $\lambda=\perp,\parallel$ vs. $\lambda=0$,
    as in Eq.~(\ref{eq:C9-truncated}), is not a good approximation. 
\end{itemize}

\section{Fit results}
\label{sect:fit}

\subsection{Extraction of $C_9$ from data}
\label{sect:C9fit}

Following the approach adopted in Ref.~\cite{Bordone:2024hui}, we have analyzed the recent experimental results in Ref.~\cite{LHCb:2025mqb}. In particular, we use the configuration (vi) of Ref.~\cite{LHCb:2025mqb},
 which partially includes muon mass effects, keeps the $CP$ asymmetries fixed at zero, uses the $S_i$ basis for the $CP$-averaged observables, and uses smaller $q^2$ bins. We analyze the data allowing for the presumed short-range component of $C_9$ to vary across different $q^2$ bins and for different helicities. More precisely, we have fitted the binned data 
 by means of the amplitude~(\ref{eq:C9matrixel}) with  the substitution 
\begin{equation}
C_9 \to C_9^{\mathrm{eff}, \lambda} (q^2)
    =   C^{\lambda,{\rm bin}}_{9}  + Y(q_0^2) +  \Delta C_9^\lambda(q^{2})_{\rm res}\,,
    \label{eq:C9fit}
\end{equation}
where $C^{\lambda,{\rm bin}}_{9}$ are the (real) fitted parameters, 
$q^2_0=-4.6~{\rm GeV}^2$, and 
where $Y(q_0^2)$ and  $\Delta C_9^\lambda(q^{2})_{\rm res}$ are given in Eq.~(\ref{eq:Yq0-fin}) and (\ref{eq:C9LD}), respectively. 
The theoretical uncertainty on $Y(q_0^2)$ in Eq.~(\ref{eq:Yq0-fin}) is not included in the fit: we treat it as an additional source of error when comparing the fitted results of $C^{\lambda,{\rm bin}}_{9}$ with the SM value of $C_9$.

The fitting procedure follows exactly the procedure adopted in Ref.~\cite{Bordone:2024hui},
to which we refer for more details. We recall here only the key points:  we extract the best-fit point by minimizing the $\chi^2$ function
\begin{equation}
    \chi^2 = \sum_{i,j} [\mathcal{O}_i^\mathrm{exp}-\mathcal{O}_i^\mathrm{theory}](V_\mathrm{theory}+V_\mathrm{exp})_{ij}^{-1} [\mathcal{O}_j^\mathrm{exp}-\mathcal{O}_j^\mathrm{theory}]
\end{equation}
where the indices $i\,,j$ run over all the observables $\mathcal{O}_{i(j)}$. The matrices $V_\mathrm{theory}$ and $V_\mathrm{exp}$ are the theoretical and experimental covariances, respectively. 
All the errors correspond to $68\%$ confidence interval, which we obtain by profiling the $\chi^2$ functions over the various fit parameters. 

\begin{figure}[t]
    \centering
\includegraphics[width=0.8\linewidth]{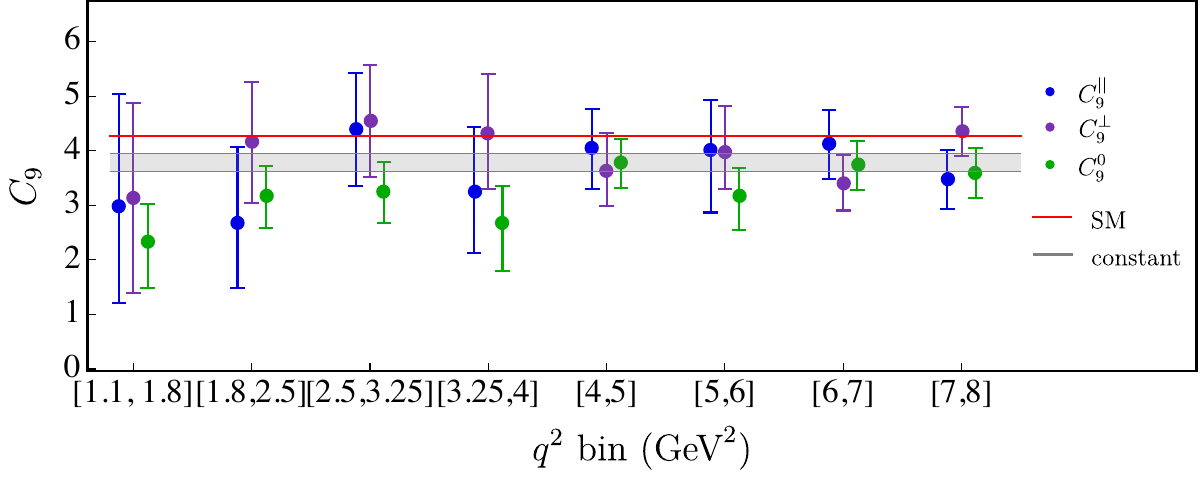}
    \caption{Fitted values of $C^{\lambda,{\rm bin}}_{9}$ in different $q^2$ bins (reported on the horizontal axis) and different helicities in the 
    low-$q^2$ region. The gray band illustrates the $1\sigma$ range of the result of the fit assuming a constant $C_9$. The SM prediction for $C_9$ is shown in red. 
    \label{fig:halfLow}}
\end{figure}

\begin{figure}[t]
    \centering
\includegraphics[width=0.8\linewidth]{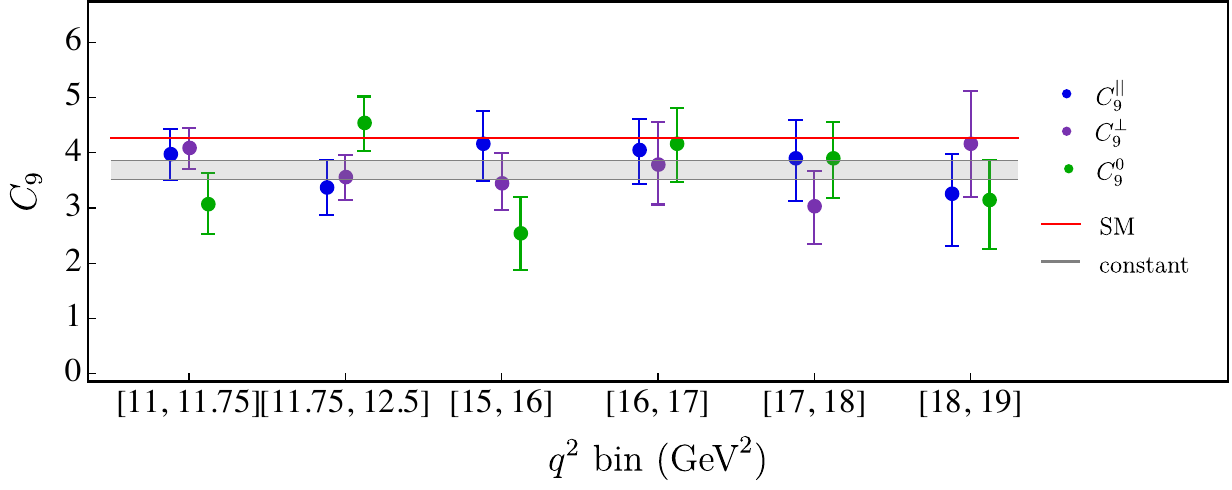}
    \caption{Fitted values of $C^{\lambda,{\rm bin}}_{9}$ in  the charmonium resonances region; notations as in Fig.~\ref{fig:halfLow}. \label{fig:halfHigh}}
\end{figure}

Most of the input parameters are treated as in Ref.~\cite{Bordone:2024hui}. In particular, we use the local factors in~\cite{Bharucha:2015bzk} (see \cite{Gubernari:2018wyi,Gubernari:2023puw} for other results).
Two notable differences are the updated resonance parameters, for which we use the results in Appendix~\ref{app:etaV}, and the value of $C^{\rm eff}_7$.
The real part of $C^{\rm eff}_7$ is treated as in Ref.~\cite{Bordone:2024hui} (see discussion at the end of Sect.~\ref{sect:nonpert4q}); however, as already stated, 
we allow a small but non-negligible value of ${\rm Im} (C^{\rm eff}_7)$. The latter is set to ${\rm Im} (C^{\rm eff}_7)=-0.05 \pm 0.02$ from an {\em a posteriori}
analysis of $S_{7,8,9}$ in the low $q^2$ region, as discussed below. 
The results of the fit are shown in Figs.~\ref{fig:halfLow} and \ref{fig:halfHigh}. Numerical results summarizing the quality of the different fits and the corresponding values of $C_9$ are reported in Table~\ref{tab:C9fit}. 

\paragraph{Determination of ${\rm Im} (C^{\rm eff}_7)$.}
As discussed in Sect.~\ref{sect:SiTH}, within our framework the low-$q^2$ values of $S_{7,8,9}$ are sensitive to ${\rm Im}(C^{\rm eff}_7)$, with $S_7$ expected to exhibit the largest effect. Current data favor negative values of $S_7$ of $O(10^{-1})$ at low $q^2$~\cite{LHCb:2025mqb}. According to Eq.~(\ref{eq:S7lowq}), this points to ${\rm Im}(C^{\rm eff}_7)\sim - ({\rm few})\times 10^{-2}$.
However, the current experimental uncertainties are still too large to allow a precise determination of ${\rm Im}(C^{\rm eff}_7)$ from data alone. On the other hand, explicit theoretical estimates indicate that ${\rm Im}(C^{\rm eff}_7)$ is expected to be at most of $O(10^{-2})$~\cite{Khodjamirian:2010vf,Altmannshofer:2014rta,Paul:2016urs}. Motivated by these considerations, we adopt the reference value ${\rm Im}(C^{\rm eff}_7)=-0.05\pm0.02$.
This figure is chosen such that, within the $1\sigma$ range, it covers both the central value of the theoretical estimate in~\cite{Paul:2016urs} ($-0.03$) and the 
the central value of ${\rm Im}(C^{\perp, \rm eff}_7+C^{\parallel \rm eff}_7)/2$ obtained by LHCb~\cite{LHCb:2024onj} ($-0.07$).
The consistency of this choice is supported by the quality of the fit in the low $q^2$ region, whose $p$-value increases from 
$p=0.25$ for ${\rm Im}(C^{\rm eff}_7)=0$ to $p=0.70$ for ${\rm Im}(C^{\rm eff}_7)=-0.05\pm0.02$. As expected, this improvement is driven primarily by the different $\chi^2$ contribution associated with $S_7$ in the two scenarios. The corresponding determinations of $\Re(C^\lambda_9)$ are fully compatible, indicating that the extraction of the short-distance contribution is largely insensitive to the assumed value of ${\rm Im}(C^{\rm eff}_7)$ within the considered range.

\begin{table}[t]
    \centering \renewcommand{\arraystretch}{1.5}
    \begin{tabular}{l|c||c|c||c}
    Type of fit & $q^2$~region~(GeV$^2$) & $p$-value & $\chi^2$/dof & Output \\ \hline \hline
      $C^{\lambda,{\rm bin}}_9$~~free & [1.1--8.0]
            & 0.55 
            &  0.96
            & see Fig.~\ref{fig:halfLow} \\ 
      $C^{\lambda,{\rm bin}}_9 = C_9$ & [1.1--8.0]
            & 0.70 
            & 0.91 
            &  $C_9 = 3.78  ^{+0.17}_{-0.16}$  \\ \hline
       $C^{\lambda,{\rm bin}}_9$~~free & [11--12.5],~[15--19] 
            & 0.24 
            & 1.14 
            & see Fig.~\ref{fig:halfHigh} \\
       $C^{\lambda,{\rm bin}}_9 = C_9$ & [11--12.5],~[15--19]   
            &  0.07        
            &  1.28
            &   $C_9=3.69 ^{+0.17}_{-0.16}$  \\  \hline
       $C^{\lambda,{\rm bin}}_9$~~free & all
            & 0.51 
            & 0.99 
            & ---
            \\
       $C^{\lambda,{\rm bin}}_9 = C_9$ & all   
            &  0.32 
            & 1.05
            &   $C_9=3.75 ^{+0.13}_{-0.12}$  \\ 
    \end{tabular}
    \caption{$p$-values, $\chi^2$/dof, and output results in different fit scenarios.}
    \label{tab:C9fit}
\end{table}

\subsection{Discussion} 

As can be seen from Fig.~\ref{fig:halfLow}, in the whole low-$q^2$ region the results are fully compatible with a $q^2$- and helicity-independent value of $C_9$, as expected if the latter is dominated by short-distance dynamics. The values extracted in individual bins and for different helicities are not identical, but their variation is fully consistent with the statistical fluctuations expected from the experimental uncertainties. Correspondingly, the fit obtained under the assumption of a constant $C_9$ exhibits an excellent quality, as shown in Table~\ref{tab:C9fit}.

In the high-$q^2$ region and in the interval between the two narrow charmonia the quality of the fit assuming a constant $C_9$ is slightly worse, but still acceptable. 
Also the fit allowing the coefficients $C^{\lambda,{\rm bin}}_9$ to vary freely does not produce a fit as good as in the low-$q^2$ region.  This can be viewed as an indication that the parameterization of rescattering terms adopted in our analysis has problems in describing perfectly the data, locally in $q^2$, in the high-$q^2$ region. 
Such a limitation is not surprising given the rapid variation of the absorptive part of the amplitude---the only component that is not allowed to vary in the fit---shown in Fig.~\ref{fig:C9LD}, together with the approximations employed in its description. In particular, the use of constant widths becomes less appropriate for the broader charmonium states. More generally, as pointed out in Ref.~\cite{Dong:2026fdi}, the region above the open-charm threshold would require a detailed treatment of coupled-channel effects that goes beyond a naive sum of Breit--Wigner amplitudes. On the other hand, the extracted value of $C_9$ (assuming a constant value) in the high-$q^2$ region is fully compatible with that obtained in the low-$q^2$ region. 

Considering all the available $q^2$ bins, the results are fully compatible with a $q^2$- and helicity-independent value of $C_9$. In other words, there is no statistical evidence of unaccounted-for long-distance contributions with $q^2$- and/or helicity-dependence. 
In view of this observation, taking into account the analyses of 
Ref.~\cite{Isidori:2024lng,Isidori:2025dkp} according to which long-distance contributions not related to single-particle intermediate states should also exhibit a 
large $q^2$-dependence, we consider the result of the global fit assuming a constant $C_9$ as the current best estimate of the short-distance value of this Wilson coefficient. The difference with the SM prediction can be quantified as follows 
\begin{equation}
\Delta C_9 =
    C^{\rm SM}_9 - C_9^{\rm fit} = 0.52 \pm 0.13_{\rm fit} \pm 0.21_{q_0} \pm 
    0.04_{C_9^{\rm SM}} = 0.52 \pm 0.25\,,
    \label{eq:DC9fit}
\end{equation}
where we used 
$C^{\rm SM}_9(\mu_b)= 4.27 \pm 0.04$~\cite{EOSAuthors:2021xpv}
with $\mu_b=4.2~{\rm GeV}$. 
The uncertainty labeled as $q_0$ in  Eq.~(\ref{eq:DC9fit})
is the one associated with the subtraction point that, by construction, takes into account the variation of $\mu_b$. 
As anticipated, the tension with the SM is around the $2\sigma$ level.\footnote{
It is worth noting that, so far, the only other channel where a similar dispersive analysis is available is the $B\to K\mu^+\mu^-$ decay. The results obtained by LHCb in this mode via a complete amplitude analysis~\cite{LHCb:2026suh} also show a reduced tension with the SM in the value $C_9$; however, a direct comparison with the result in Eq.~(\ref{eq:DC9fit}) is not straightforward given the overall quality of the fit in the $B\to K$ case is not as good as for $B\to K^*$, especially if $C_{10}$ is fixed to its SM value, as implied by $\mathcal{B}(B_s\to\mu^+\mu^-)$. This problem might be related to an underestimate of the uncertainty on the $B\to K$ form  factors~\cite{Ciuchini:2026hxy}.}

\subsection{Postdictions of $S_5$ and $S_{7,8,9}$}
\label{sect:PostSi}

\begin{figure}[t]
    \centering
\includegraphics[width=0.65\linewidth]{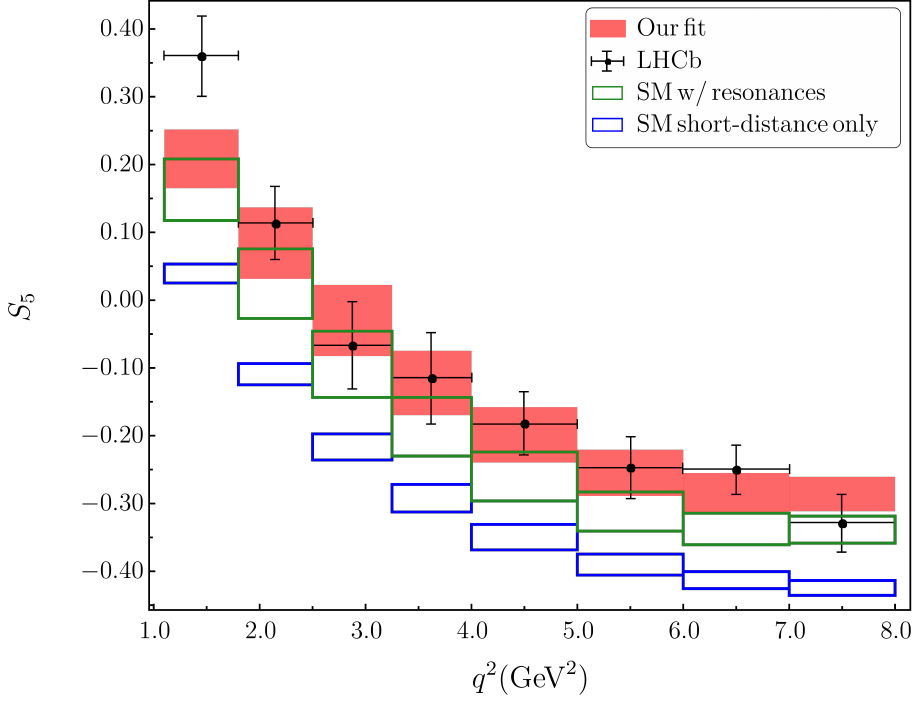} 
    \caption{Predictions of $S_{5}$ vs.~experimental data (black dots). The red bands indicate our postdiction using the short-distance value of $C_9$ from the global fit and including all long-distance corrections. The green bands denote the result using $C_{9}^{\rm SM}$ and including all long-distance corrections.  The blue bands denote the short-distance SM contribution, i.e.~the~result obtained using $C_{9}^{\rm SM}$ and neglecting all non-local matrix-element corrections (including those to the dipole operator).
    \label{fig:S5} }
\end{figure}

\begin{figure}[t]
    \centering
    \hspace{-5pt}
\includegraphics[width=0.48\linewidth]{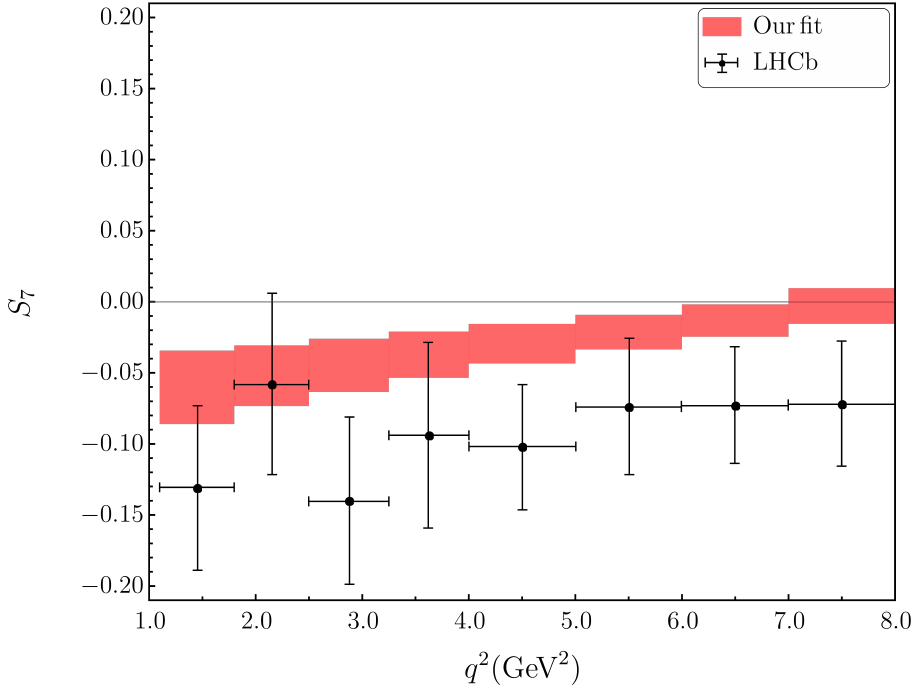} 
    \hspace{5pt} 
\includegraphics[width=0.48\linewidth]{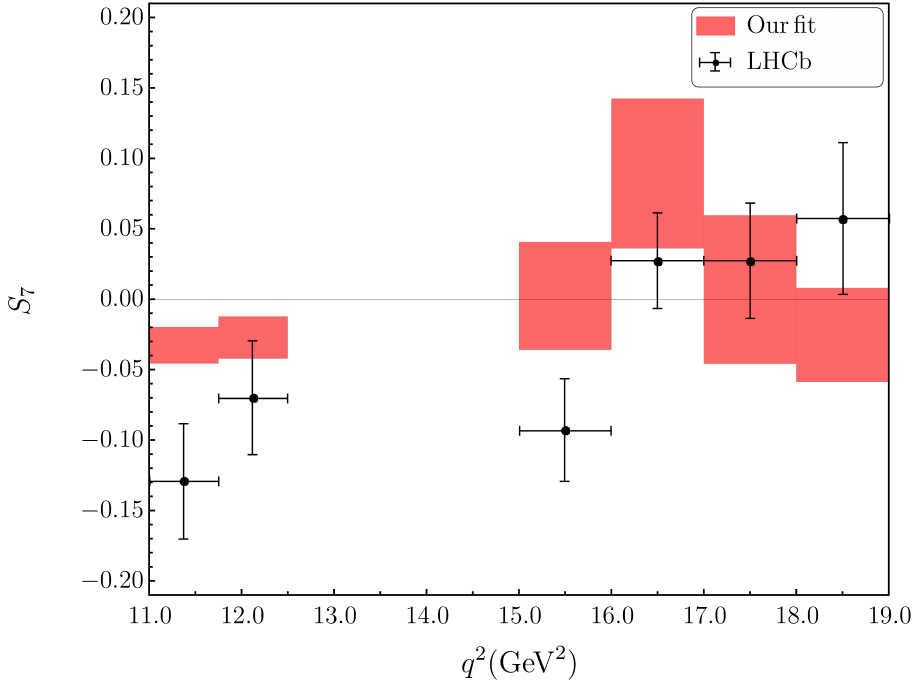}  \\[8pt]
    \hspace{-5pt}
\includegraphics[width=0.48\linewidth]{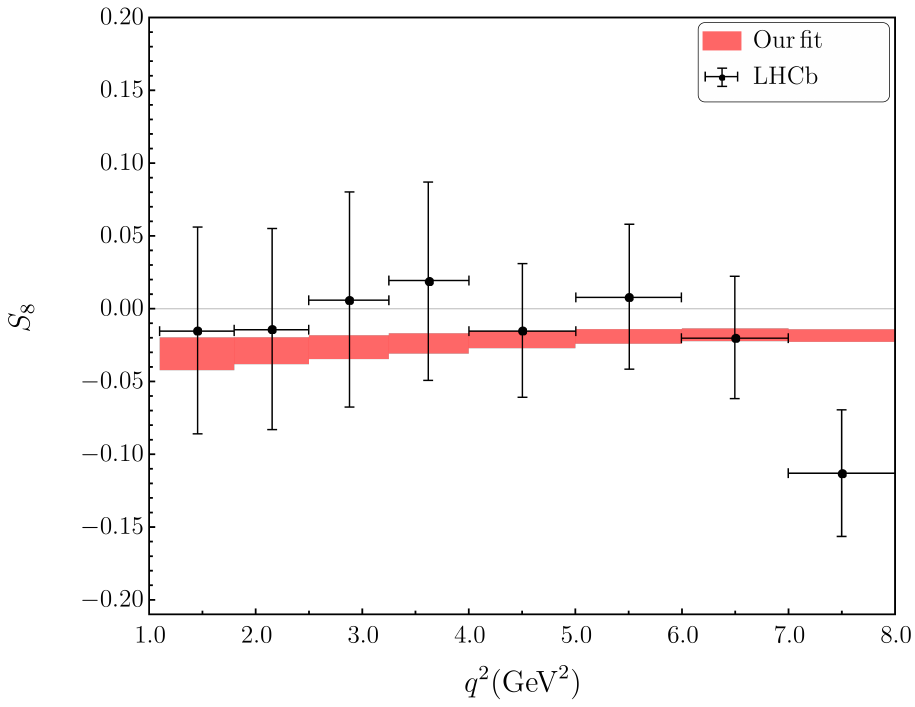}  
    \hspace{5pt}
\includegraphics[width=0.48\linewidth]{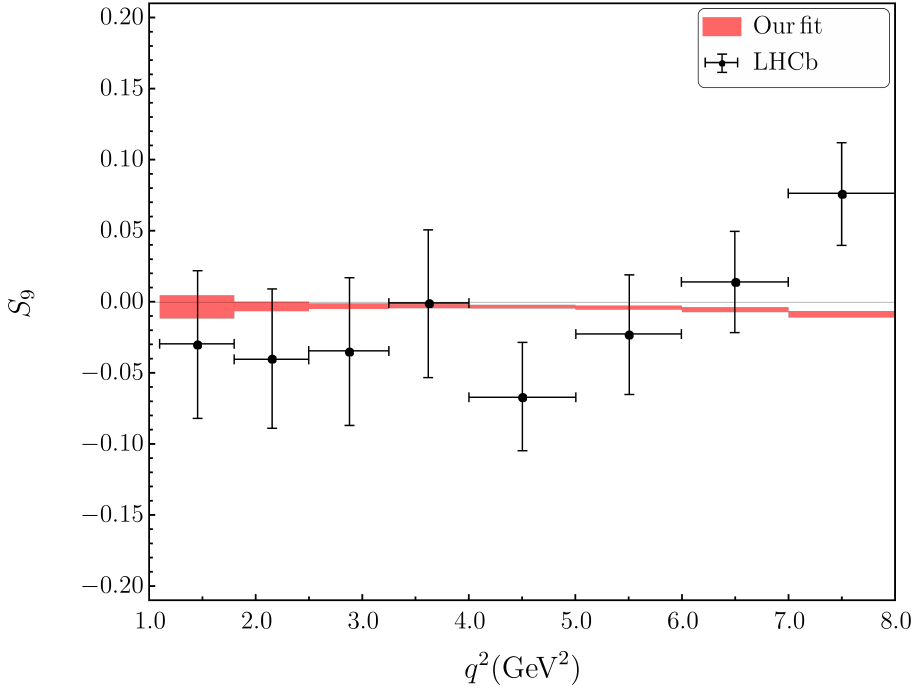}  
    \caption{Postdiction of $S_{7}, S_8, S_9$ vs.~experimental data (black points) in different $q^2$ ranges. 
    The upper panel shows  $S_{7}$ in the low-$q^2$ (left) and high-$q^2$ (right) regions. The lower panel shows  $S_8$ (left) and $S_9$  in the low-$q^2$ region. 
    In all plots the red bands indicate our postdiction using the short-distance value of $C_9$ from the global fit and including all long-distance corrections.
    \label{fig:Si} }
\end{figure}

In Figs.~\ref{fig:S5} and \ref{fig:Si} we show the postdictions for $S_5$ and $S_{7,8,9}$ obtained from our fit assuming a constant $C_9$. The $S_5$ distribution has historically played a key role in signaling a deviation from the SM prediction possibly due to a non-standard value of $C_9$ (see e.g.~Refs.~\cite{Capdevila:2023yhq,Altmannshofer:2026cwk}).
By contrast, as discussed in Sect.~\ref{sect:SiTH}, the observables  $S_{7,8,9}$ provide direct probes of rescattering effects, since they are proportional to the absorptive parts of the decay amplitudes.
In other words, $S_5$ is very sensitive to the short-distance value of $C_9$ and the dispersive component of the rescattering amplitude, while $S_{7,8,9}$ are largely insensitive to the value of $C_9$ and very sensitive to the absorptive part of the rescattering amplitude. The comparison of these observables with our predictions is therefore particularly informative. 

Although we refer to the red bands in Figs.~\ref{fig:S5}--\ref{fig:Si}  as  ``postdictions'', it is worth stressing that they retain a high predictive power. Indeed, once the resonance parameters are fixed by data (from a $q^2$ region which is largely excluded by our fit), the only quantities that are not determined \emph{a priori} are ${\rm Re}(C_9)$, extracted from the fit, and  ${\rm Im}(C^{\rm eff}_7)$, whose value is fixed using both theoretical considerations and the very first bin of $S_7$ (as discussed in Sect.~\ref{sect:C9fit}). The resulting predictions are therefore a direct consequence of the highly constrained structure of the dispersive description of the non-local matrix elements. 

 Fig.~\ref{fig:S5} illustrates the anatomy of the different effects that, when combined, allow us to obtain an excellent description of current data for $S_5$. 
 At very low values of $q^2$, the largest correction to the short-distance contribution is played by the non-local corrections to the dipole term, i.e.~the shift from $C_7$ to ${\rm Re}(C^{\rm eff}_7)$ 
in Eq.~(\ref{eq:C7eff}). For higher $q^2$ values,
the non-local corrections to $C^{\lambda,\rm eff}_9$ are those playing the leading role.  In the whole range of $q^2$, the small shift in the short-distance value of $C_9$ leads to an overall improvement in the quality of the fit. 

Concerning $S_{7,8,9}$, it is interesting to note that the postdictions in Fig.~\ref{fig:Si}
are significantly more precise than currently available experimental data in the whole low-$q^2$ region.  This is a consequence of the constrained form of the absorptive part of the amplitude in Eq.~(\ref{eq:C9LD}). Given such accuracy in predicting 
$S_{7,8,9}$, future more precise experimental results on these distributions could provide a decisive test for the description of the absorptive part of the amplitude and, more generally, of the 
dispersive description of the non-local matrix elements. 

The second important point to note about 
$S_{7,8,9}$  is that only $S_{7}$ should be clearly different from zero according to the parameterization in Eq.~(\ref{eq:C9LD}), and to the coefficients of the resonances determined by experiments. 
This expectation is supported by current data, as shown in  Fig.~\ref{fig:Si}.

\subsection{Comparison with other approaches} 
It is instructive to compare the present analysis with alternative strategies that have recently been adopted to describe the $B\to K^*\mu^+\mu^-$ amplitude and determine the value of $C_9$ from data.  

\begin{itemize}
\item{} {\em Data-driven analysis with resonances~\cite{Bordone:2024hui}.} Beside a series of minor improvements (most notably the inclusion of light resonances and wider charmonium states, and the improved determination of the subtraction point), the current analysis shares most methodological aspects with Ref.~\cite{Bordone:2024hui}. The significant difference in the extraction of $C_9$ ($|\Delta C_9|$ reduces by about $50\%$ in the current fit) is mainly due to the different values of the resonance phases.  
In Ref.~\cite{Bordone:2024hui} these were not determined by data but naively extrapolated from the $B\to K \ell^+\ell^-$ case. As shown in Fig.~\ref{fig:C9LD}, the phases determined by data imply a negative correction to $C_9^{\lambda, {\rm eff}}$, for all polarizations, that in turn leads to a smaller correction to $C_9$.

\item{} {\em Amplitude fit by LHCb~\cite{LHCb:2024onj}.}
Our determination of $\Delta C_9$ is fully compatible, within uncertainties, with that obtained in~\cite{LHCb:2024onj}. This is not surprising given that the two analyses adopt the same functional form for the long-distance contributions and both lead to good overall fits.
However, we stress that this agreement is a non-trivial 
{\em a posteriori} consistency check, given that the two analyses are structurally different. In all our fits, the leading resonance regions (both at very low $q^2$ and close to the $J/\Psi$ and $\Psi(2S)$ peaks) are excluded. The resonance parameters are included as 
external inputs and varied within their uncertainties, rather than being determined 
from the fit itself as in~\cite{LHCb:2024onj}.
As a result,  the presence of additional long-distance contributions, if any,  would be signaled by the variation of  $C_9^{\lambda,{\rm eff}}$ in different $q^2$ bins and/or different helicities and/or wider $q^2$ regions (e.g.~low vs.~high $q^2$). The fact that this does not happen, at the current level of precision, is a non-trivial consistency check of the approach. 

\item{} {\em Polynomial parameterization of non-local matrix elements~\cite{Altmannshofer:2026cwk,Ciuchini:2026hxy}.} 
Fitting non-local matrix elements assuming a generic polynomial form in $q^2$  precludes, by construction, a determination of $C_9$ from data. The approach is certainly conservative, and retains predictive power in  SM tests not involving the vector-current component of the amplitude; however, it gives up in trying to determine $C_9$ given the unavoidable degeneracy between short- and (uncostrained) long-distance terms~\cite{Ciuchini:2026hxy}.
A further limitation of that approach is the fact that it can only deal with a limited portion of the spectrum.  
\\
Our approach, using experimental information over the entire spectrum,
aims at extracting both the short- and long-distance contributions simultaneously. As such, it is more predictive.
The trade-off is the introduction of a certain degree of model dependence in the way non-local terms are parameterized. However, as we have shown, it also offers a series of consistency checks that can be used, a posteriori, to validate the parameterization. 
\\
Within our approach we can quantitatively address the question raised in Ref.~\cite{Altmannshofer:2026cwk} if the evidence of non-vanishing  $S_7$ helps solve the apparent discrepancy between data and SM prediction in $S_5$. The answer obtained with current data is given by Fig.~\ref{fig:S5} and Fig.~\ref{fig:Si}: our analysis indicates that the currently observed non-vanishing values of $S_7$ are compatible with the long-distance corrections predicted within the dispersive framework. These corrections significantly improve the description of $S_5$, but do not completely remove the residual tension in the determination of $C_9$, which remains approximately at the $2\sigma$ level.   
\item{} {\em Explicit estimates of non-local matrix elements~\cite{Hurth:2025vfx}.} The explicit estimates of non-local matrix elements at low $q^2$ based on $1/m_b$, $1/E_{K^*}$ and $1/m_c$ expansions
\cite{Beneke:2004dp,Khodjamirian:2012rm}, employed in the recent data analysis of Ref.~\cite{Hurth:2025vfx} with suitable conservative errors, tend to underestimate long-distance effects compared to the dispersive approach. This is why the central values of $|\Delta C_9|$ determined in Ref.~\cite{Hurth:2025vfx} are larger than those that we obtain (see Tab.~\ref{tab:C9fit}). The underestimate of long-distance effects grows with growing $q^2$, and this explains why in Ref.~\cite{Hurth:2025vfx}  the central value of $|\Delta C_9|$ increases when the $q^2\in[6,8]~{\rm GeV}^2$ bins are also included. This should be contrasted with our analysis, where the determination of $|\Delta C_9|$ is very stable over the entire $q^2$ spectrum and is particularly precise and consistent in the whole region  
$q^2\in[2.5,8]~{\rm GeV}^2$ (see Fig.~\ref{fig:halfLow}).
\end{itemize}

\section{Conclusions}
\label{sect:Conc} 

We have presented an updated data-driven analysis of short- and long-distance dynamics in the decay $B\to K^{*}\mu^+\mu^-$. The analysis exploits the recent LHCb measurements of the full angular distribution, performed in finely spaced $q^2$ bins, together with the corresponding global amplitude fit. The improved experimental information provides a substantially more stringent test of non-local hadronic effects. The new data provide clear evidence for rescattering effects, not only through the detailed $q^2$ dependence of the decay amplitudes, but also via the angular observable $S_7$, which offers a particularly clean probe of the absorptive part of the amplitude. This constitutes an important qualitative step forward with respect to previous analyses, where the evidence for rescattering originated mainly from the distortions of the differential spectrum while approaching the resonance region.

Within the dispersive framework adopted in this work, the observed rescattering effects in the entire low-$q^2$ region, up to 
$q^2\approx 8~{\rm GeV}^2$, are well described by the contribution of vector-meson resonances, whose parameters are determined independently from the global amplitude analysis. In this region, the values of $C_9$ extracted independently in different $q^2$ bins and for different helicity amplitudes are fully compatible with a helicity- and $q^2$-independent contribution, as naturally expected for a short-distance Wilson coefficient. Current data therefore provide a non-trivial validation of the dispersive description and allow a significantly more reliable extraction of the short-distance component of the decay amplitude.

The situation is slightly more involved above the $\psi(2S)$ resonance. In this region the resonance-based parameterization is not as good as at low $q^2$, signaling the limitation of describing rescattering effects only through a sum of Breit--Wigner amplitudes. This behaviour is not unexpected, given the onset of open-charm thresholds and the increasing importance of coupled-channel dynamics. Nevertheless, after averaging over the whole high-$q^2$ region, the extracted value of $C_9$ remains fully compatible with that determined below the $J/\psi$ resonance, indicating no statistically significant evidence for additional helicity- or $q^2$-dependent short-distance contributions.

The resulting determination of the short-distance value of $C_9$ 
shows that long-distance effects inferred from data substantially reduce the discrepancy with the SM compared to analyses where these contributions are not constrained by experimental information. At the same time, our analysis does not provide compelling evidence that rescattering effects alone can account for the long-standing tension in the determination of $C_9$. Within the present framework, and with current data, a residual discrepancy at the level of approximately $2\sigma$ remains.

A further important outcome of this work concerns the observables $S_7$, $S_8$, and $S_9$. Owing to the constrained structure of the absorptive part of the amplitudes within the dispersive framework, we obtain precise postdictions for these quantities, together with a sum rule that links them and is largely insensitive to the details of the rescattering parameterization. The predicted uncertainties are significantly smaller than those of the current measurements, making these observables particularly promising probes of non-local dynamics. Future experimental determinations of $S_7$, $S_8$, and $S_9$ will therefore provide powerful tests of the dispersive description and valuable internal consistency checks of the data.

The present analysis illustrates how current, and especially future, precision measurements can provide deep insights into the structure of non-local hadronic effects in exclusive $b\to s\ell^+\ell^-$ transitions. The combination of analyticity, dispersion relations, and perturbative constraints on the one hand, with precise binned measurements and global amplitude fits on the other, provides a powerful framework to disentangle short- and long-distance dynamics in $B\to V\ell^+\ell^-$ decays. The level of precision that can ultimately be achieved remains an open question, but the results presented here indicate that this is a promising strategy to pursue. In this respect, while waiting for more precise measurements of $B\to K^*\mu^+\mu^-$, it 
would be particularly interesting to perform analogous studies of other channels, such as $B_s\to\phi\mu^+\mu^-$. 
The combined information from these complementary channels, as well as the comparison between exclusive and inclusive modes~\cite{Alvarez-Cartelle:2026nhp}, may provide one of the most powerful tools to unravel the interplay between weak interactions and non-perturbative QCD in rare $B$ decays.

\section*{Acknowledgments}
We thank Wolfgang Altmannshofer for useful discussions and cross-checks about the expressions of $S_{7,8,9}$, and Ulrik Egede, Konstantinos Petridis, Eluned Smith and Mark Smith for useful clarifications about the LHCb analyses. The work of M.B. is supported by the Cluster of Excellence \textit{PRISMA}$^{++}$ (EXC 2118/2) funded by the German Research Foundation (DFG) under Germany’s Excellence Strategy (Project ID 390831469).
The work by G.I. and A.T.~is supported by the Swiss National Science Foundation, projects No. PCEFP2-194272 and 2000-1-240011.

\appendix

\section{Determination of the resonance parameters}
\label{app:etaV}

The $\kappa_\lambda(q^2)$ introduced in Eq.~\ref{eq:Al-norm} to express the helicity amplitudes, normalised as in 
Eq.~(\ref{eq:Al-norm}),
are 
\begin{align}
\kappa_\perp (q^2)  &=  \sqrt{2} N (q^2)\,   \frac{  \llambda(q^2)^{1/2} }{m_B+m_{K^*}  } \,,  \\
\kappa_\parallel (q^2) &=   -  \sqrt{2}  N(q^2)\,  (m_B+m_{K^*} )\,,  \\
\kappa_0 (q^2)  &=-  8 N (q^2)\,   \frac{  m_B  m_{K^*} }{      \sqrt{q^2} } \,.
\end{align}
where 
%
%
\begin{align}
N(q^2)  =  |\cN| \sqrt{ \frac{ q^2  }{  384 \pi^3  m_B^3}   \llambda^{1/2}(q^2) }\,.
\end{align}
Taking into account the modification  $ C_9 \to C_9^{\mathrm{eff}, \lambda} (q^2)$ and 
considering only the resonant terms in $C_9^{\mathrm{eff}, \lambda} (q^2)$, we can extract 
the contribution given by each resonance  to the decay $B\to K^* \ell^+\ell^-$ in a 
 specific helicity final state:
\begin{align}
\frac{ d\Gamma_V^\lambda}{dq^2}  = 2    |\kappa_\lambda(q^2)|^2   (16\pi^2 )^2  |\eta_V^\lambda|^2
\left| \frac{(q^2-q_0^2)}{(m_V^2-q_0^2)}  A_V^\text{res}(q^2) \right|^2  \,.
\end{align}
The subtraction point is  irrelevant to determine $\Gamma_V^\lambda$ in the narrow width approximation,
where
\begin{align}
 | A_V^\text{res}(q^2) |^2  \, \to  \, \pi m_V \Gamma_V\, \delta (q^2-m_V^2) \,.
\end{align}
Given the definition of the $A_V^\lambda$ coefficients by LHCb in~\cite{LHCb:2024onj},
\begin{align}
|(A_V^\lambda)_{\rm exp}|^2 =  \cB( B\to K^* V^\lambda) \cB( V\to e^+e^-)\,,
\end{align}
we get 
\begin{align}
\Gamma_V^\lambda \equiv \Gamma_B \,  | (A_V^\lambda)_{\rm exp} |^2  = 2^9 \pi^5  m_V \Gamma_V   |\kappa_\lambda(M_V^2)|^2   |\eta_V^\lambda|^2\,,
\end{align}
which  allows us to derive 
\begin{align}
 |\eta_V^\lambda| =  \left[  \frac{ \Gamma_B } { 2^9 \pi^5 m_V \Gamma_V  |\kappa_\lambda(m_V^2)|^2 } \right]^{1/2}    \times  | (A_V^\lambda)_{\rm exp} |\,.
 \end{align}
Separating each helicity, we obtain
\begin{align}
 |\eta_V^\perp | &=   \frac{1}{G_F \alpha    |V_{tb} V_{ts}^*| } \left[ \frac{  3  \Gamma_B }{   \Gamma_V} \,
 	\frac{  m_B^3 (m_B+m_{K^*})^2} {   m^3_V  \llambda^{3/2}(m_V^2)   } \right]^{1/2}    \times  | (A_V^\perp)_{\rm exp} |\,, 
    \label{eq:eta-det1}
    \\
|\eta_V^\parallel | &=   \frac{1}{G_F \alpha    |V_{tb} V_{ts}^*| } \left[   \frac{  3 \Gamma_B } {   \Gamma_V } \, 
    	\frac{m_B^3 }{  m^3_V   (m_B+m_{K^*} )^2   \llambda^{1/2}(m_V^2)  } \right]^{1/2}   \times  | (A_V^\parallel)_{\rm exp} |\,,  
    \label{eq:eta-det2}
        \\
 |\eta_V^0 | &=  \frac{1}{G_F \alpha    |V_{tb} V_{ts}^*| } \left[   \frac{3 \Gamma_B}{32  \Gamma_V}\,  
   	\frac{ m_B    } { m_V m^2_{K^*}  \llambda^{1/2}(m_V^2) } \right]^{1/2}   \times  | (A_V^0)_{\rm exp} |\,.
    \label{eq:eta-det3}
     \end{align}

For the charmonium states, all the $A_V^\lambda$ and the corresponding phases are reported in the LHCb paper~\cite{LHCb:2024onj}.
The only exception is  $A_{J/\Psi}^0$, which can be deduced from the PDG 
value of  $\cB(B\to J/\psi K^*)$~~\cite{ParticleDataGroup:2024cfk}.
The $\eta^\lambda_V$ computed from the
$A_V^\lambda$ using Eqs.~(\ref{eq:eta-det1})--(\ref{eq:eta-det3}) are
 reported in Table~\ref{Tab:eta_cc}.

\begin{table}[t]
    \centering
    \begin{tabular}{c|c|c}
      resonance &  $M$ (GeV) & $\Gamma$ (GeV) \\
      \hline
      $J/\Psi$ & 3.0969 & $92.6 \times 10^{-6}$ \\
      $\Psi(2S)$ &  3.6861 & $294.0 \times 10^{-6}$ \\
     $ \Psi(3770)$ &  3.7737 & 0.0272 \\
      $ \Psi(4040)$  & 4.04 & 0.084 \\
      $\Psi(4160)$  &  4.191 & 0.069 
    \end{tabular}
 \vskip 0.5 true cm 
    \centering
    \begin{tabular}{c|c|c|c}
     resonance & $\eta_\perp$ & $\eta_\parallel$ & $\eta_0$  \\
     \hline
      $J/\Psi$ & $25.309\pm 0.923$ & $11.375 \pm 0.430$ & $12.894 \pm   0.613$ \\
       $\Psi(2S)$ & $3.9858 \pm  0.3216$ & $1.4065 \pm 0.1271$ & $1.6726 \pm  0.1461$ \\
       $ \Psi(3770)$ & $(1.9414 \pm 0.8183) \times 10^{-2}$ & $(7.1328 \pm 2.5917) \times 10^{-3}$ & $(4.7359 \pm 2.0605) \times 10^{-3}$ \\
       $ \Psi(4040)$ & $(8.9976 \pm 7.0012) \times 10^{-3}$ & $(2.2928 \pm 1.6858) \times 10^{-3}$ & $(4.9517 \pm 1.7348) \times 10^{-3}$ \\
       $\Psi(4160)$ & $(2.0713 \pm 1.3487) \times 10^{-2}$ & $(2.1414 \pm 2.7510) \times 10^{-3}$ & $(2.1189 \pm 2.0885) \times 10^{-3}$
    \end{tabular}
\vskip 0.5 true cm 
    \begin{tabular}{c|c|c|c}
     resonance & $\delta_\perp$ & $\delta_\parallel$ & $\delta_0$ \\
     \hline
        $J/\Psi$ 
        &   $-2.13\pm 0.06$ 
        &   $-1.69\pm 0.06$ 
        &   $-1.92\pm 0.054$\\
        $\Psi(2S)$  
        &  $-2.98\pm 0.21$ 
        &  $-1.70\pm 0.26$ 
        &  $-2.54 \pm 0.18$ \\
        $ \Psi(3770)$ & $-2.90\pm 0.45$ & $0.66 \pm 0.38$ & $2.65 \pm 0.46$  \\
       $ \Psi(4040)$ & $2.97\pm0.83$ & $-2.85\pm0.84$ & $2.91\pm0.36$  \\
       $\Psi(4160)$ & $3.03	\pm 0.10$ & $1.55\pm 0.95$ & $-1.02\pm0.95$
    \end{tabular}
    \caption{Masses, widths, $\eta_V^\lambda$ and  $\delta_V^\lambda$ for the charmonium resonances
    \label{Tab:eta_cc}}
\end{table}

For the light resonances, we compute the 
$A_V^\lambda$ using  input values from PDG~\cite{ParticleDataGroup:2024cfk}
and Ref.~\cite{LHCb:2018hsm,LHCb:2014xzf}
(for the $B\to VK^*$ decays). These inputs are
summarized in Table~\ref{Tab:lightRes}.
Using theses results in Eqs.~(\ref{eq:eta-det1})--(\ref{eq:eta-det1}) we obtain the $\eta_V^\lambda$ reported in
Table~\ref{Tab:eta_LR}. The values of the 
phases are deduced using the phase differences 
$\delta_V^\lambda-\delta_V^0$ from~\cite{LHCb:2018hsm,LHCb:2014xzf}
and the values of $\delta_V^0$ in~\cite{LHCb:2024onj}.

\begin{table}[t]
    \centering
    \begin{tabular}{c|c|c|c|c|}
      resonance &  $M$ (GeV) & $\Gamma$ (GeV) & $\cB(V\to e^+e^-)$ \\
      \hline
         $\rho$ & 0.770 & $0.144$ & $(4.72\pm 0.05)\times 10^{-5}$ \\
         $\omega$ & 0.782 & $8.68 \times 10^{-3}$ & $(7.41 \pm 0.19)\times 10^{-5}$ \\
        $\phi$ & 1.019 & $4.25 \times 10^{-3}$ & $(2.96 \pm 0.033) \times 10^{-4}$\\
    \end{tabular}
    \vskip 0.5 true cm
    \centering
    \begin{tabular}{c|c|c|c|c}
      resonance &    $\cB(B \to V K^*)$ & $f_V^\perp$ &   $f_V^\parallel$ &$f_V^0$  
                \\ \hline
        $\rho$ &  $(3.9\pm1.3)\times 10^{-6}$ &  $0.401\pm0.040$ & $0.435\pm0.045$ & $0.164\pm0.027$ 
                \\
        $\omega$ &  $(2.0\pm0.6)\times 10^{-6}$   & $0.100\pm0.130$ & $0.220\pm0.200$ & $0.680\pm0.230$  
                \\
        $\phi$ &  $(1.00\pm0.05)\times 10^{-5}$ & $0.221\pm0.021$ & $0.282\pm0.022$ & $0.497\pm0.024$ 
    \end{tabular}
    \caption{Masses, widths, and decay rates of the light resonances  
    \label{Tab:lightRes}}
    \end{table}

\begin{table}[t]    
    \centering
    \begin{tabular}{c|c|c|c} 
     resonance & $\eta_\perp$ & $\eta_\parallel$ & $\eta_0$  \\
     \hline
      $\rho$ &  $0.0057\pm 0.0010$ &  $0.0041 \pm 0.0007$ &  $0.00045\pm0.00008$  \\
      $\omega$ & $0.0102\pm0.0066$  & $0.0105\pm0.0051$ & $0.0033\pm0.0007$    \\
      $\phi$ & $0.0671\pm0.0036$  &  $0.0516\pm0.0025$  &  $0.0161\pm0.0006$ 
    \end{tabular} 
\vskip 0.5 true cm
    \begin{tabular}{c|c|c|c}
     resonance & $\delta_\perp$ & $\delta_\parallel$ & $\delta_0$  \\
     \hline
      $\rho$ &  $-2.56\pm  0.86$ &  $0.61 \pm 0.86$ &  $1.38\pm0.88$ \\
      $\omega$ & $1.97\pm1.50$ & $-1.46\pm1.38$ & $-0.49\pm1.52$    \\
      $\phi$ & $2.73\pm1.13$  &  $2.66\pm1.13$  &  $0.10\pm1.13$ 
    \end{tabular}
    \caption{Values of $\eta_V^\lambda$ and  $\delta_V^\lambda$ for the light resonances
        \label{Tab:eta_LR}}
\end{table}

\newpage

\bibliographystyle{JHEP}
\bibliography{references.bib}

\end{document}